\documentclass[aps,prl,reprint,amsmath,amssymb,superscriptaddress,longbibliography,nofootinbib]{revtex4-2}

\usepackage[T1]{fontenc}
\usepackage[utf8]{inputenc}
\usepackage{graphicx}
\usepackage{xcolor}
\usepackage{comment}
\usepackage{amsthm}
\usepackage{mathtools}

\newtheorem{thm}{Theorem}
\newtheorem{lemma}[thm]{Lemma}

\newcommand{\pazocal}{\mathcal}
\newcommand{\HH}{\pazocal{H}}
\newcommand{\DD}{\pazocal{D}}
\newcommand{\XX}{\pazocal{X}}
\newcommand{\fament}{\hat{E}_d}
\newcommand{\locc}{\mathrm{LOCC}}

\DeclareMathOperator{\Tr}{Tr}
\newcommand{\norm}[1]{\left\|#1\right\|}
\DeclarePairedDelimiter{\ceil}{\lceil}{\rceil}

\newcommand{\ket}[1]{\lvert #1\rangle}
\newcommand{\bra}[1]{\langle #1\rvert}
\newcommand{\ketbra}[1]{\ket{#1}\!\bra{#1}}

\newcommand{\bb}{\begin{equation}\begin{aligned}}
\newcommand{\ee}{\end{aligned}\end{equation}}
\newcommand{\bbb}{\begin{equation*}\begin{aligned}}
\newcommand{\eee}{\end{aligned}\end{equation*}}

\newcommand{\prlheading}[1]{\par\medskip\noindent\textit{\textbf{#1.}}\textbf{---}\ignorespaces}

\usepackage[hidelinks]{hyperref}
\hypersetup{
    pdftitle={Universal entanglement distillation},
    pdfauthor={Salvatore Tirone, Francesco Anna Mele, Vittorio Giovannetti, Ludovico Lami}
}

\begin{document}

\title{Universal entanglement distillation}

\author{Salvatore Tirone}
\email{tironesalvatore@gmail.com}
\affiliation{QuSoft, Science Park 123, 1098 XG Amsterdam, the Netherlands}
\affiliation{Korteweg--de Vries Institute for Mathematics, University of Amsterdam, Science Park 105--107, 1098 XG Amsterdam, the Netherlands}

\author{Francesco Anna Mele}
\email{fmele@caltech.edu}
\affiliation{Institute for Quantum Information and Matter, California Institute of Technology, Pasadena, California 91125, USA}

\author{Vittorio Giovannetti}
\email{vittorio.giovannetti@sns.it}
\affiliation{NEST, Scuola Normale Superiore and Istituto Nanoscienze, Consiglio Nazionale delle Ricerche, Piazza dei Cavalieri 7, 56126 Pisa, Italy}

\author{Ludovico Lami}
\email{ludovico.lami@gmail.com}
\affiliation{Scuola Normale Superiore, Piazza dei Cavalieri 7, 56126 Pisa, Italy}

\begin{abstract}
%We present an entanglement distillation protocol that distills entanglement from copies of an unknown state in an optimal way. The asymptotic Bell-pair rate of our protocol is exactly equal to the optimal rate one would obtain if the unknown state were known, which in turn is equal to the distillable entanglement of the unknown state. Our approach combines LOCC tomography with an adaptive distillation routine, utilizing a novel robustness lemma to ensure that finite-sample estimation errors do not compromise asymptotic purity. We also extend this framework to characterize worst-case distillation over families of states with partial structural knowledge. These results resolve a long-standing open question and provide technical tools widely applicable to other quantum resource theories.
We present a universal entanglement distillation protocol that asymptotically achieves the distillable entanglement of any unknown bipartite state in a fixed finite dimension. That is, its ebit yield attains every rate below the optimum that would be available if the state were known, with vanishing trace-distance error. Our approach combines tomography by local operations and classical communication with an adaptive distillation routine, using a local robustness lemma to control finite-sample estimation errors. We also characterize worst-case distillation from a family of states: the optimal uniform rate is the minimum distillable entanglement on the family's closure. These results show that ignorance of the state need not incur an asymptotic rate penalty.
\end{abstract}

\maketitle

\prlheading{Introduction}
Entanglement is one of the most valuable resources for quantum technologies. Quantum communication, error correction, cryptography, and tests of nonlocality~\cite{teleportation,dense-coding,Bennett-error-correction,Ekert91,RennerPhD,Brunner-review} rely on entanglement, often supplied in the form of Bell pairs, or ebits. A crucial task in quantum information science is therefore \textit{entanglement distillation}~\cite{Bennett-distillation,Bennett-distillation-mixed}, which aims to extract approximate ebits from many identical copies of a bipartite state using only local operations and classical communication (LOCC)~\cite{LOCC,gap_sep_locc}.
The usual formulation of entanglement distillation assumes that Alice and Bob have a complete classical description of their shared state. In practice, this description may not be available, and finite-sample tomography cannot, in general, determine an arbitrary unknown state exactly. In this work, we therefore study entanglement distillation when the state is partially or entirely unknown. Related state-agnostic questions have been considered in settings ranging from entanglement concentration~\cite{Matsumoto2007,Hayashi2017,hayashi2001variablelengthuniversalentanglement}, state merging and entanglement distillation~\cite{BjelakovicBocheJanssen2013,BocheJanssen2014,LamiRegulaTakagi2026}, and quantum communication~\cite{BjelakovicBocheNoetzel2009,hayashi_univ_coding,Hayashi_wiretap,hayashi_alphabet} to work extraction~\cite{safranek-unknown-work,watanabe2025universalworkextractionquantum,lumbreras2025quantumstateagnosticworkextraction,black_box_watanabe,canzio2025extractingchargingenergyunknown,faist2026,Guarnieri2026WorkExtraction}.

We prove the existence of an optimal universal protocol for entanglement distillation from unknown bipartite states.
Our approach starts with an LOCC tomography phase, followed by a distillation routine selected on the basis of the resulting estimate.
Remarkably, this two-stage strategy asymptotically achieves the full distillable entanglement $E_d$ of every input state with vanishing error.
Here, the local dimension is known and the input consists of identical, independently prepared copies; it is the state itself that is unknown.
Thus, prior classical knowledge of the state is not a prerequisite for optimal entanglement distillation.

Central to this finding is a technical lemma establishing the local robustness of entanglement distillation.
For any distillable state $\rho$ and any rate $0 < R < E_d(\rho)$, we show that there is a sufficiently small open ball around $\rho$ on which a single sequence of LOCC maps achieves rate $R$, with error tending to zero uniformly over the ball.
The radius may depend on both $\rho$ and $R$.
Leveraging this robustness, we also show that the worst-case distillable entanglement over a nonempty family of states $\XX$ equals the minimum of $E_d$ on its topological closure $\bar{\XX}$.

Conceptually, this framework connects entanglement distillation with quantum learning theory~\cite{Aaronson2007,Aaronson2018,Huang2020,AnshuArunachalam2024}, and in particular with quantum state tomography~\cite{Haah2016,ODonnell2016-1,ODonnell2016-2, PelecanosSpileckiTangWright2025}. While quantum learning theory often focuses on the sample complexity of state characterization, here we use its statistical guarantees to select a resource-distillation protocol. The key point is physically intuitive: the state need not be learned exactly; instead, a sufficiently accurate estimate suffices to identify a robust protocol, and the fraction of copies spent on learning can vanish asymptotically.

Our result thus answers whether a permanent asymptotic rate penalty is an unavoidable toll for ignorance.
Under the assumptions above, it is not.
The robustness lemma may also be useful in other resource theories, provided that analogous local distillation guarantees can be established.

\prlheading{Preliminaries} Let us briefly review the setting and the technical tools required for our work. We consider finite-dimensional complex Hilbert spaces, with $\HH_A \cong \HH_B \cong \HH$ and $\HH = \mathbb{C}^d$, where $d \geq 2$ is the local dimension and $A$ and $B$ are held by Alice and Bob, respectively. Given a Hilbert space $\HH$, we denote by $\DD(\HH)$ the corresponding set of quantum states, namely the positive semidefinite operators on $\HH$ with trace one. A shared state $\rho_{AB}$ thus belongs to $\DD(\HH_A \otimes \HH_B)$; we omit subsystem labels when this does not give rise to any ambiguity. All logarithms are to base two.

We measure the distance between states $\rho$ and $\sigma$ by the trace distance $\frac{1}{2}\norm{\rho - \sigma}_1$ and 
%use the fidelity convention 
the fidelity $F(\rho,\sigma) \coloneqq \norm{\sqrt{\rho}\sqrt{\sigma}}_1$~\cite{Uhlmann-fidelity}.
We write $B_r(\rho) \coloneqq \left\{\sigma\in \DD(\HH):\, \frac12\|\rho-\sigma\|_1<r\right\}$ for the open trace-distance ball of radius $r$ around $\rho$, relative to the state space.
Trace distance and fidelity obey the Fuchs--van de Graaf inequalities~\cite{Fuchs1999}
\bb
1 - F(\rho,\sigma)
\leq \frac{1}{2}\norm{\rho - \sigma}_1
\leq \sqrt{1 - F(\rho,\sigma)^2}.
\ee

We now turn to entangled states and entanglement distillation.
For a start, a \textit{separable} state is a density operator of the form $\rho = \sum_j p_j \rho_j^{(A)} \otimes \rho_j^{(B)}$, where $\{p_j\}$ is a probability distribution and the $\rho_j^{(A)}$ and $\rho_j^{(B)}$ are local states~\cite{Werner}. A bipartite state is \textit{entangled} if it is not separable. %A Bell pair 
An ebit is the two-qubit state $\Phi_2 = \ketbra{\Phi_2}$, with $\ket{\Phi_2} \coloneqq (\ket{00} + \ket{11})/\sqrt{2}$.
The goal of entanglement distillation is to transform $n$ copies of $\rho$ into $\ceil{Rn}$ approximate %Bell pairs 
ebits using only LOCC. We would like the coefficient $R$, referred to as the \textit{rate}, to be as large as possible.
The \textit{distillable entanglement} $E_d(\rho)$ is the supremum of all rates $R \geq 0$ for which there exists a sequence of LOCC maps whose output on $\rho^{\otimes n}$ approaches $\Phi_2^{\otimes \ceil{Rn}}$ in trace distance as $n \to \infty$.
A state $\rho$ is \textit{distillable} if $E_d(\rho) > 0$.

Importantly, in the above definition the sequence of LOCC maps realizing the distillation process is allowed to depend on $\rho$. This models the assumption that a complete classical description of $\rho$ is available, a possibly unrealistic assumption. Our goal, instead, is to construct a single universal sequence, independent of the input state, that attains every rate below $E_d(\rho)$ on each $\rho$. The rules of the game are that 
%In this definition, the sequence of maps may depend on $\rho$. Our first goal is instead to find a single universal sequence, chosen independently of the input state, that attains every rate below $E_d(\rho)$ on each $\rho$.
the protocol may adapt to its measurement outcomes, but not to an externally supplied description of the state. We allow its output to contain a classically announced, possibly random number of approximate %Bell pairs
ebits. The error is measured by the trace distance from ideal %Bell pairs
ebits of the announced number, averaged over that number; the rate guarantee also requires that the probability of producing too few pairs tends to zero. Theorem~\ref{thm:univ_rate} makes these guarantees mathematically precise.

\begin{figure}[t]
    \centering
    \includegraphics[width=\linewidth]{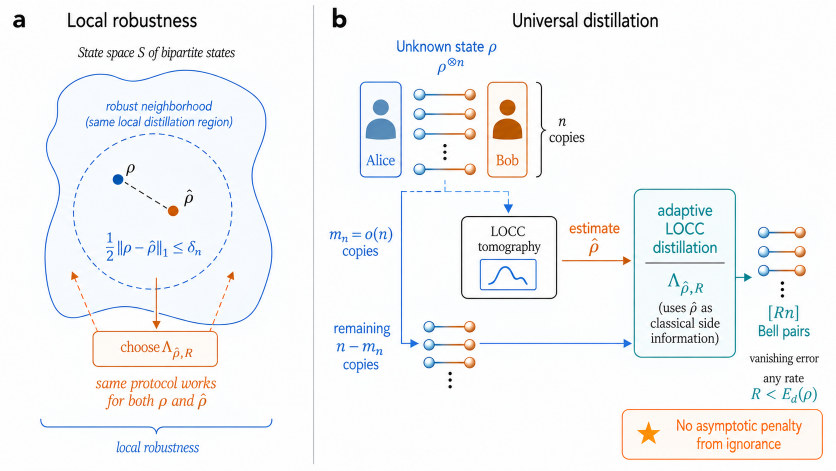}
    \caption{Universal entanglement distillation.
    (a) Local robustness (Lemma~\ref{lemma:robustness}): for any distillable state $\rho$ and any $0 < R < E_d(\rho)$, a single sequence of LOCC maps achieves rate $R$ with uniformly vanishing error on a sufficiently small ball around $\rho$.
    (b) A sublinear number of input copies is used for LOCC tomography. Then, the tomographic estimate is used to select a robust distillation routine for the remaining copies, yielding asymptotically optimal universal distillation.}
    \label{fig:protocol}
\end{figure}

We also consider the setting in which Alice and Bob know that their state belongs to a given nonempty family $\XX \subseteq \DD(\HH_A \otimes \HH_B)$.
We can define the \textit{universal distillable entanglement} for the family $\XX$ as the following rate:

\begin{multline}
    \fament(\XX) \coloneqq \sup\Bigl\{R \geq 0 :
    \lim_{n\to\infty}\inf_{\Lambda_n}\sup_{\rho\in\XX} \\
    \frac12\left\|\Lambda_n(\rho^{\otimes n})-
    \Phi_2^{\otimes\ceil{Rn}}\right\|_1=0\Bigr\}.
\label{def:universaldist}
\end{multline}
%Here $A^0$ and $B^0$ are local qubit systems. 
Here, the infimum runs over all LOCCs $\Lambda_n$ with input system given by $n$ copies of $AB$ and output system composed of $\ceil{Rn}$ pairs of qubits, shared between Alice and Bob. Intuitively, $\fament(\XX)$ quantifies the supremum of rates at which a universal protocol can distill %Bell pairs 
ebits from all states $\rho\in \XX$ with vanishing error. For every choice of a universal protocol $\Lambda_n$, the error is computed on the worst-case input state, which explains why the supremum over $\rho\in \XX$ is inside the infimum over $\Lambda_n$. Therefore, this figure of merit takes into account Alice and Bob's ignorance of the state, since they only know that it is contained in $\XX$. The rate $R=0$ corresponds to an empty output.

\prlheading{LOCC tomography} We start by designing a tomography protocol for bipartite states that uses only LOCC. It combines two ingredients. First, for each copy of the unknown bipartite state, Alice sends a noisy version of her subsystem to Bob by means of a \textit{classical teleportation} procedure: Alice and Bob apply the usual quantum teleportation protocol~\cite{teleportation}, but with the maximally entangled resource $\Phi_d=d^{-1}\sum_{i,j=0}^{d-1}\ket{ii}\bra{jj}$ replaced by the separable isotropic state
$W_d \coloneqq \frac{1}{d+1}\Phi_d+
(1-\frac{1}{d+1})\frac{\mathbb{I}}{d^2}$~\cite{Horodecki1999}. After this LOCC procedure, Bob holds a perturbed copy of the input state, namely,\footnote{A little thought reveals that the above procedure is equivalent to Alice measuring her system with a two-design pure-state POVM $\big(\psi\, \mathrm{d}\mu(\psi)\big)$, communicating the outcome ``$\psi$'' to Bob, and Bob preparing an auxiliary system in the state $\psi$.}
\bb
    \mathcal{N}(\rho)
    \coloneqq
    \frac{1}{d+1}\rho
    +
    \left(1-\frac{1}{d+1}\right)
    \frac{\mathbb{I}_A}{d}\otimes\Tr_A[\rho].
\ee
Second, Bob performs global mixed-state tomography on the resulting copies of
$\mathcal{N}(\rho)$. We use the standard fact that, for an unknown state acting
on a $D$-dimensional Hilbert space, the optimal number of copies required to
learn the state within trace-distance error $\varepsilon$ and with constant
failure probability is of order
$D^2/\varepsilon^2$~\cite{Haah2016,ODonnell2016-1,ODonnell2016-2}. More
recently, Ref.~\cite{PelecanosSpileckiTangWright2025} gave a particularly
simple protocol with sample complexity
$O((D^2+\log(1/\delta))/\varepsilon^2)$, which is also optimal with respect to
the dependence on the failure probability
$\delta$~\cite{ScharnhorstSpileckiWright2025}.
Finally, Bob classically applies the inverse map
$\mathcal{N}^{-1}(\Theta)=(d+1)\Theta-\mathbb{I}_A\otimes\Tr_A[\Theta]$ to the
tomographic estimate and post-processes the result into a valid density
operator. Since here $D=d^2$ and the inversion of $\mathcal{N}$ amplifies the
trace-distance error by a factor $O(d)$, the overall sample complexity for learning bipartite states via LOCC is
$O(d^6/\varepsilon^2)$ for constant failure probability. More precisely, we have the following.

\begin{thm}
\label{thm:estimate}
For all $\varepsilon,\delta\in(0,1)$, there exists an LOCC tomography protocol
that, given
\bb
    N
    =
    O\left(
        \frac{d^6+d^2\log(1/\delta)}{\varepsilon^2}
    \right)
\ee
copies of an unknown bipartite state
$\rho\in\DD(\mathbb{C}^{d}\otimes\mathbb{C}^{d})$, outputs an estimator
$\widehat{\rho}\in\DD(\mathbb{C}^{d}\otimes\mathbb{C}^{d})$ such that $\frac12\left\|\widehat{\rho}-\rho\right\|_1\leq\varepsilon$ with probability at least $1-\delta$.
\end{thm}

The full proof of the above theorem is given in the supplemental material~\cite{SM}. Recently, the authors of~\cite{leone2025entanglement} introduced an LOCC tomography protocol for pure bipartite states with sample complexity $\tilde{\Theta}(d^2/\varepsilon^2)$. Our algorithm has worse scaling in $d$, but applies to arbitrary mixed states.

\prlheading{Universal distillation}  The main result of this paper is a consequence of the following Lemma:

\begin{lemma}[Local robustness] \label{lemma:robustness}
For any distillable state $\rho$, and for any \(0<R<E_d(\rho)\), there exist a radius \(r>0\) and a sequence of LOCC maps \(\Lambda_n^{(\rho,R)}\) such that
\begin{equation}
    \lim_{n\to\infty}
    \sup_{\sigma\in B_r(\rho)}
    \frac12
    \norm{\Lambda_n^{(\rho,R)}(\sigma^{\otimes n})-
    \Phi_2^{\otimes\ceil{Rn}}}_1
    =0 \; .
\label{eq:local_robustness}
\end{equation}
\end{lemma}
In simple terms, the Lemma above guarantees that, given a distillable state $\rho$ and any rate $0<R<E_d(\rho)$, there exists a region of distillable states and a fixed LOCC protocol, such that, for any state in this region, the fixed algorithm allows us to distill at rate $R$. The core content of this robustness result is that the fixed protocol is not sensitive to a slight change of the input state. 
%In other words, the distillation protocol is robust with respect to local perturbations of the state. 
%
A %Corollary 
simple consequence of the above lemma is that the distillable entanglement is lower semicontinuous everywhere, which, in turn, also entails that the set of distillable states is open in trace-distance topology for any $d$.

Now, we %fulfill the main goal of our work: 
present the main result of our work: namely, we exhibit an algorithm which, given a generic bipartite state $\rho$, is able to asymptotically distill 
%Bell pairs 
ebits at optimal rate $E_d(\rho)$ with vanishing error. A pictorial representation of this protocol is depicted in Figure~\ref{fig:protocol}. The existence and the performance of this protocol are formalized as follows.
%in the following Theorem.

\begin{thm}[Universal entanglement distillation] \label{thm:univ_rate}
%There exists a single sequence of LOCC protocols, independent of the input state and target rate, with the following property: for any bipartite quantum state\(\rho\in\DD(\mathbb C^d\otimes\mathbb C^d)\), and any rate \(0<R<E_d(\rho)\), the universal protocol applied to \(\rho^{\otimes n}\) asymptotically distills at least $\ceil{Rn}$ %Bell pairs \ludo{ebits} with probability tending to one and with vanishing error in trace distance, averaged over the announced output length. The averaged error also vanishes when $E_d(\rho)=0$. 
%\ludo{[To be looked at again: shouldn't the protocol depend on some slack parameter $\delta>0$, and achieve rate $\big(E_d(\rho) - \delta\big)_+$ for all $\rho$? In the present form we are formally saying that the protocol distills at asymptotic rate \emph{equal} to $E_d(\rho)$ for all $\rho$, which seems too strong? (Indeed, consider the asymptotic rate of the fixed protocol on $\rho$, which is a number $F(\rho)$ depending only on $\rho$. The statement says that $F(\rho) \geq R$ for all $R < E_d(\rho)$, so $F(\rho) \geq E_d(\rho)$.)]} 
There exists a state-agnostic sequence of LOCC protocols with the following property: for any bipartite quantum state $\rho \in \DD(\mathbb C^d \otimes \mathbb C^d)$, and any rate $0<R<E_d(\rho)$, the universal protocol applied to $\rho^{\otimes n}$ asymptotically distills $\ceil{Rn}$ ebits with probability tending to one and with vanishing error in trace distance.
\end{thm}

The proof of this result is given in great detail in Ref.~\cite{SM}. Here, we give a sketch of the universal protocol. 
First, Alice and Bob share $n$ copies of the state $\rho$ and choose a sublinear number $m_n$ of them to obtain a classical estimate of the original state. For every positive rational rate $q$, they can find a countable open cover of the set of distillable states with distillable entanglement strictly larger than $q$, called $\mathcal{O}_q$, and they can fix a suitable $\delta(n)$ precision parameter. At blocklength $n$, only a finite menu of these balls is available, growing slowly enough that tomography uses $o(n)$ copies and the distillation error vanishes uniformly over the menu. 
After that, they employ the tomographic primitive described in Theorem~\ref{thm:estimate} to obtain a classical estimate $\hat{\rho}_n$ of the true state $\rho$, such that $\frac12\left\| \hat{\rho}_n  - \rho \right\|_1 \le \delta(n)$ with probability tending to one. A ball is selected only if its robustness neighborhood contains the entire closed $\delta(n)$-ball around $\hat{\rho}_n$, and hence also $\rho$ on tomographic success. 
The balls are chosen to be small enough to fulfill the requirements of Lemma~\ref{lemma:robustness}. Therefore, since $\hat{\rho}_n$ and $\rho$ are close enough, Alice and Bob can choose an LOCC distilling entanglement at the largest admissible rate $q$ for $\hat{\rho}_n$ with vanishing error. Thus, due to Lemma~\ref{lemma:robustness}, the same operation distills at the same rate for $\rho$. For every fixed $0<R<E_d(\rho)$, a suitable ball with a fixed rate $q>R$ eventually becomes available with probability tending to one. So, they perform the aforementioned LOCC on the remaining $n-m_n$ copies to asymptotically distill at least $\ceil{Rn}$ %Bell pairs 
ebits with vanishing error; the strict rate slack pays for the $o(n)$ copies used for tomography. 

\prlheading{Worst-case family distillation} Up to this point we have discussed universal distillation for unknown single states. It is also useful to consider the case where Alice and Bob have prior partial knowledge on their shared state, meaning that they know that it belongs to a given nonempty family $\XX \subseteq \DD(\pazocal{H}_A\otimes\pazocal{H}_B)$. In such a scenario, we are interested in the worst-case quantity $\fament(\XX)$. A similar construction to that used in Theorem~\ref{thm:univ_rate} allows us to prove another result on worst-case distillation formalized in the following statement.

\begin{thm}[Worst-case distillation] \label{thm:fament}
For any nonempty family of states $\XX\subseteq\DD(\mathbb{C}^d\otimes\mathbb{C}^d)$ the universal distillable entanglement is
\bb
    \fament(\XX) = \min_{\rho\in\bar{\XX}}E_d(\rho) \; ,
\ee
where $\bar{\XX}$ is the closure of $\XX$.
\end{thm}

In Ref.~\cite{SM} we give a rigorous proof of the above statement. Here, we provide a sketch of the demonstration. We have proved that the distillable entanglement is lower semicontinuous everywhere; therefore it attains its minimum on a compact set, so $\inf_{\rho\in\bar{\XX}}E_d(\rho) = \min_{\rho\in\bar{\XX}}E_d(\rho)$. Moreover, in~\cite{SM} we prove that $\fament(\XX)=\fament(\bar{\XX})$. So, since the set of quantum states in finite dimension $d$ is compact with respect to the trace-norm, any closed family of states is a compact set. Therefore, for any $0<R<\min_{\rho\in\bar{\XX}}E_d(\rho)$, we can use an open cover of the family $\bar{\XX}$ by smaller balls inside robustness neighborhoods at a common rate $q$ strictly between $R$ and this minimum, similar to the one used in Theorem~\ref{thm:univ_rate}. By compactness of $\bar{\XX}$ the cover can be made finite, giving a uniform tomographic tolerance and enough rate slack to pay for tomography. Then, we employ the same universal protocol used above to prove that, for any rate $R < \min_{\rho\in\bar{\XX}}E_d(\rho)$, there exists a universal sequence of LOCC maps $\{\Lambda_n\}_{n\in\mathbb{N}}$ such that

\bb 
\lim_{n\to\infty}\sup_{\rho\in\bar{\XX}}\frac12\left\| \Lambda_n\left(\rho^{\otimes n}\right) - \Phi_2^{\otimes \ceil{Rn}} \right\|_1 = 0 \; .
\ee 
This proves that $\fament(\XX) \ge \min_{\rho\in\bar{\XX}}E_d(\rho)$; when the minimum is zero, this inequality is immediate. The reverse inequality is an easy consequence of the definition of the universal distillable entanglement.

\prlheading{Discussion} We have given a non-constructive existence proof for a universal entanglement distillation protocol for unknown bipartite mixed states, demonstrating that it is possible to achieve both the optimal distillation rate and a vanishing asymptotic error without a priori state knowledge. By integrating a novel LOCC-based quantum tomography procedure, which utilizes only a vanishing fraction of the available copies, with a state-adaptive distillation routine, this result establishes that the fundamental limits of entanglement distillation for known states are attainable without a priori knowledge of the state. 

Our findings resolve a major open question in the field, confirming that the existence of a universal protocol for optimal-rate, vanishing-error distillation is not hindered by the lack of initial state information. 
The key technical ingredient enabling this advancement is a novel robustness lemma for distillation protocols. This lemma guarantees that minor deviations in the classical state estimate, which naturally arise from the finite-sample LOCC tomography stage, do not compromise the asymptotic optimality or the purity of the final distilled %Bell pairs.
ebits.

Beyond the fully agnostic setting, we have also obtained a rigorous result on the worst-case distillation over families of states, characterizing scenarios where we possess partial knowledge of the state. This result bridges the gap between fully state-dependent routines and completely agnostic protocols, providing optimal operational bounds when the input state is known to belong to a specific set or satisfies certain symmetry constraints.

On a conceptual level, our framework formalizes an intrinsic synergy between quantum learning theory and operational resource manipulation. By embedding statistical learning guarantees directly into a distillation pipeline, we demonstrate how data-acquisition costs can be accommodated when processing quantum resources. This integration establishes a rigorous precedent for analyzing quantum resource theories under epistemic uncertainty, treating information gathering and resource extraction as an interconnected task.

Looking forward, a highly promising perspective is the application of our technical tools to other quantum resource theories. In particular, the core mathematical machinery developed here may be generalized to find similar universal results in quantum communication. For instance, these tools could be leveraged to design universal quantum error-correcting codes or to achieve optimal transmission rates over completely unknown quantum channels without prior channel characterization. Additionally, determining the optimal sample complexity of LOCC tomography remains an open task to ascertain whether our current upper bound is tight, as is the extension of this universal framework to multipartite systems. Ultimately, our protocol establishes a rigorous baseline and opens a new chapter in quantum information theory where information-theoretic bounds are reached via self-consistent and learning-assisted protocols.

\prlheading{Acknowledgements} We are grateful to Andreas Winter for many insightful comments on an early version of this work, which was presented at the conference ``Quantum Resources 2026'' (Tokyo, 16--20 March 2026). FAM and LL acknowledge financial support from the European Union (ERC StG ETQO, Grant Agreement no.\ 101165230).

\emph{AI statement.---} The authors began this project in late 2022 and completed a first version in 2025 without using generative AI. ChatGPT 5.5 was later used to improve the main result by refining one of the proofs suitably adapting our human-generated arguments. The authors carefully verified all AI-assisted refinements and take full responsibility for the content of this work.

\emph{Note added.---}  During the final stage of writing this manuscript, we became aware of an independent work by Takagi et al.~\cite{ryuji_univ}. Here, the authors show that universal distillation is possible through state merging for compound sources.  

% References cited only in the Supplemental Material.
%\nocite{Horodecki-unified,Mele2024introductiontohaar,Vollbrecht2003,devetak2005}

\bibliography{biblio}

\end{document}

\title{\texorpdfstring{Universal Entanglement Distillation \\[1ex] Supplemental Material}{Universal entanglement distillation: Supplemental material}}

\author{Salvatore Tirone}
\email{tironesalvatore@gmail.com}
%\affiliation{Scuola Normale Superiore, I-56126 Pisa, Italy}
\affiliation{QuSoft, Science Park 123, 1098 XG Amsterdam, the Netherlands}
\affiliation{Korteweg--de Vries Institute for Mathematics, University of Amsterdam, Science Park 105-107, 1098 XG Amsterdam, the Netherlands}

\author{Francesco Anna Mele}
\email{francesco.mele@sns.it}
\affiliation{NEST, Scuola Normale Superiore and Istituto Nanoscienze, Piazza dei Cavalieri 7, IT-56126 Pisa, Italy}

\author{Vittorio Giovannetti}
\email{vittorio.giovannetti@sns.it}
\affiliation{NEST, Scuola Normale Superiore and Istituto Nanoscienze, Consiglio Nazionale delle Ricerche, Piazza dei Cavalieri 7, IT-56126 Pisa, Italy}

\author{Ludovico Lami}
\email{ludovico.lami@gmail.com}
\affiliation{Scuola Normale Superiore, Piazza dei Cavalieri 7, 56126 Pisa, Italy}

\maketitle

\tableofcontents

\section{Preliminaries}

Throughout this supplemental material, all logarithms are in base two.

Quantum systems are indicated by capital letters $A$, $B$, etc. We reserve $A_0,B_0$ for single qubits. The set of \deff{density operators} on a quantum system with Hilbert space $\HH$, i.e.\ the convex set of positive semi-definite trace class operators on $\HH$ with trace $1$, will be denoted by $\DD(\HH)$. Physically realisable transformations between quantum systems $A$ (input) and $B$ (output), denoted with capital Greek letters such as $\Lambda$, $\Gamma$, etc., are mathematically modelled by \deff{quantum channels}, i.e.\ completely positive trace-preserving linear maps $\TT(\HH_A)\to \TT(\HH_B)$, where $\TT(\HH)$ stands for the Banach space of trace-class operators on $\HH$.

Distance between two quantum states $\rho,\rho'$ can be measured by the \deff{trace distance} $\frac12 \|\rho-\rho'\|_1$, where $\|X\|_1\coloneqq \Tr |X| = \Tr \sqrt{X^\dag X}$ is the trace norm, or alternatively via the \deff{fidelity}~\cite{Uhlmann-fidelity}
\bb
F(\rho,\rho') \coloneqq \left\|\sqrt{\rho}\sqrt{\rho'}\right\|_1\, .
\ee
Note that the fidelity $F(\rho,\rho')\in [0,1]$ is a measure of closeness, in the sense that the larger it is, the closer $\rho$ and $\rho'$ are --- indeed, $F(\rho,\rho')=1$ if and only if $\rho=\rho'$. Trace distance and (in)fidelity are universally comparable, in the sense that the \deff{Fuchs--van de Graaf inequalities}~\cite{Fuchs1999}
\bb
1 - F(\rho,{\rho'}) \leq \frac12 \|\rho-{\rho'}\|_1 \leq \sqrt{1-F^2(\rho,{\rho'})}
\label{Fuchs_van_de_Graaf}
\ee
hold for all $\rho,\rho'\in \DD(\HH)$. For a state $\rho$ and $r>0$, we write
\[
    B_r(\rho)
    \coloneqq
    \left\{\sigma\in\DD(\HH):\frac12\|\sigma-\rho\|_1<r\right\}
\]
for the open trace-distance ball of radius $r$ centered at $\rho$. In spite of this equivalence, the fidelity is a very useful quantity to work with, and sometimes a handier one than the trace distance. For example, unlike the latter it is multiplicative across tensor products, i.e.
\bb
F\left(\rho^{\otimes n},{\rho'}^{\otimes n}\right) = F(\rho,\rho')^n
\label{fidelity_multiplicativity}
\ee
for all $\rho,\rho'\in \DD(\HH)$.

The maximally entangled state of a two-qubit system, a.k.a.\ \deff{ebit}, is given by
\bb
\Phi_2\coloneqq \ketbra{\Phi_2}\, ,\qquad \ket{\Phi_2}\coloneqq \frac{1}{\sqrt2} \left(\ket{00} + \ket{11}\right) .
\label{ebit}
\ee
An immediate generalisation is the \deff{maximally entangled state} in dimension $d$, given by
\bb
\Phi_d\coloneqq \ketbra{\Phi_d}\, ,\qquad \ket{\Phi_d}\coloneqq \frac{1}{\sqrt{d}} \sum_{j=0}^{d-1} \ket{jj}\, .
\label{maximally_entangled}
\ee
The above state is part of the more general family of \deff{isotropic states}~\cite{Horodecki1999}, defined as convex combinations of $\Phi_d$ and the normalized projector onto its orthogonal complement
\bb
\tau_d \coloneqq \frac{\id - \Phi_d}{d^2 - 1}\, .
\label{tau}
\ee
Isotropic states are precisely those that are invariant under the \deff{twirling} quantum channel
\bb
\TT(X) \coloneqq \int \dd U\, (U\otimes U^*)\, X\, (U\otimes U^*)^\dag\, .
\label{twirling}
\ee

The joint quantum system with components $A$ and $B$ will be indicated by $AB$. The corresponding Hilbert space is $\HH_{AB} = \HH_A\otimes \HH_B$. A composite quantum system composed of $n$ copies of $A$ will instead be denoted by $A^n$. Given two bipartite systems $AB$ and $A'B'$, the set of quantum channels $AB \to A'B'$ that can be implemented with local operations and classical communication (\deff{LOCC})~\cite{LOCC} will be indicated by $\locc(AB\to A'B')$. 
\begin{comment}
Given two states $\rho_{AB},\omega_{A'B'}$, the fidelity of transformation $\rho\to\omega$ achievable by LOCC is given by
\bb
F_\locc\left(\rho_{AB}\to \omega_{A'B'}\right) \coloneqq \sup_{\Lambda\in \locc(AB\to A'B')} F\left(\Lambda(\rho_{AB}), \omega_{A'B'}\right) \; .
\ee
Analogously we can define the trace-distance of transformation $\rho_{AB}\to\omega_{A'B'}$ as

\bb
    T_{\locc}(\rho_{AB}\to\omega) \coloneqq \sup_{\Lambda\in\locc(AB\to A'B')}\frac12 \left\| \Lambda(\rho_{AB}) - \omega_{A'B'} \right\|_1 \; .
\ee
\end{comment}

Given a bipartite state $\rho_{AB}$, the \deff{distillable entanglement of $\rho_{AB}$ at error threshold} $\e\in [0,1)$ is defined to be~\cite{Bennett-distillation, Bennett-distillation-mixed}
\bb
E_d^\e(\rho_{AB}) \coloneqq \sup\left\{ R>0:\, \limsup_{n\to\infty} \inf_{\Lambda_n\in \locc\left(A^nB^n\to A_0^{\ceil{Rn}}B_0^{\ceil{Rn}}\right)} \frac12 \left\|\Lambda_n\left(\rho^{\otimes n}\right) - \Phi_2^{\otimes \ceil{Rn}}\right\|_1\leq \e\right\} \; .
\label{distillable_error_threshold}
\ee
Here and below, the supremum of an empty set of positive rates is taken to be zero.
The \deff{distillable entanglement} of $\rho_{AB}$ and the \deff{strong converse distillable entanglement} of $\rho_{AB}$ are then given by
\bb
E_d(\rho_{AB}) \coloneqq E_d^0(\rho_{AB})\, ,\qquad E_d^\dag(\rho_{AB}) \coloneqq \sup_{\e\in [0,1)} E_d^\e(\rho_{AB})\, .
\label{distillable}
\ee

While easily computable upper bounds on the distillable entanglement are relatively abundant, it is much more difficult to find matching lower bounds. One exception to this state of affairs is the \deff{hashing bound}, conjectured in general in~\cite{Horodecki-unified} based on the hashing protocol~\cite{Bennett-error-correction, Vollbrecht2003}, and subsequently proved by Devetak and Winter in a celebrated paper~\cite[Theorem~10]{devetak2005}. It reads
\bb
E_d(\rho_{AB}) \geq I(A\rangle B)_{\rho_{AB}} \coloneqq S(\rho_B) - S(\rho_{AB})\, .
\label{hashing}
\ee

\section{Two elementary estimates}

We first state two useful results that will be used repeatedly in the following.

\begin{lemma}[Tensor powers are locally Lipschitz]
\label{lem:tensor_lipschitz}
For all states \(\rho,\sigma\) and all \(k\in\mathbb N\), 
\[
    \norm{\rho^{\otimes k}-\sigma^{\otimes k}}_1
    \le
    k\norm{\rho-\sigma}_1 .
\]
\end{lemma}

\begin{proof}
Use the telescopic identity
\[
\rho^{\otimes k}-\sigma^{\otimes k}
=
\sum_{j=0}^{k-1}
\rho^{\otimes(k-j-1)}\otimes(\rho-\sigma)\otimes\sigma^{\otimes j} .
\]
The trace norm is multiplicative under tensor products, and every density operator has trace norm one. Therefore
\[
\norm{\rho^{\otimes k}-\sigma^{\otimes k}}_1
\le
\sum_{j=0}^{k-1}\norm{\rho-\sigma}_1
=
k\norm{\rho-\sigma}_1 ,
\]
completing the proof.
%Dividing by two proves the claim.
\end{proof}

\begin{lemma}[Finite-block continuity for a fixed LOCC map]
\label{lem:fixed_map_continuity}
Let \(\Gamma\in\locc\) be fixed. Let \(k,\ell\in\mathbb N\), and let \(\Phi=\Phi_2^{\otimes\ell}\). If
\[
    \Tr\!\left[\Phi\,\Gamma(\rho^{\otimes k})\right]\ge 1-\eta,
\]
then every \(\sigma\) satisfying
\[
    \half\norm{\sigma-\rho}_1\le r
\]
also satisfies
\[
    \Tr\!\left[\Phi\,\Gamma(\sigma^{\otimes k})\right]
    \ge
    1-\eta-kr .
\]
\end{lemma}

\begin{proof}
The operator \(\Phi\) is a projector, hence \(0\le\Phi\le\id\). For any two states \(\omega,\omega'\), the variational characterization of trace distance gives
\[
    \left|\Tr[\Phi(\omega-\omega')]\right|
    \le
    \half\norm{\omega-\omega'}_1 .
\]
Apply this to
\(\omega=\Gamma(\rho^{\otimes k})\) and
\(\omega'=\Gamma(\sigma^{\otimes k})\). Since trace distance cannot increase under a quantum channel,
\[
\begin{aligned}
\left|\Tr\!\left[\Phi\,\Gamma(\rho^{\otimes k})\right] - \Tr\!\left[\Phi\,\Gamma(\sigma^{\otimes k})\right]\right| 
&\le
\half\norm{\Gamma(\rho^{\otimes k})-\Gamma(\sigma^{\otimes k})}_1 \\
&\le
\half\norm{\rho^{\otimes k}-\sigma^{\otimes k}}_1 \\
&\le
k\,\half\norm{\rho-\sigma}_1 \\
&\le kr,
\end{aligned}
\]
where the third line uses Lemma~\ref{lem:tensor_lipschitz}. This proves the claim.
\end{proof}

\section{Classical teleportation and tomography of bipartite states via LOCCs}
\label{sec:classical_teleportation_locc_tomography}
In this section, we analyze tomography of arbitrary bipartite states under LOCC operations. Recently, Ref.~\cite{leone2025entanglement} introduced an LOCC tomography protocol tailored to \emph{pure} bipartite states; our result instead applies to arbitrary mixed states. The idea of our LOCC tomography protocol is simple: first, we use a version of the quantum teleportation protocol~\cite{teleportation} that can be implemented by LOCC, at the price of adding noise to the teleported state; then, we apply global quantum state tomography.

We start by briefly reviewing basic results on tomography of mixed qudit states under arbitrary global operations. For an unknown state acting on a $D$-dimensional Hilbert space, the optimal number of copies required to learn the state within trace-distance error $\varepsilon$ and with constant success probability is of order $D^2/\varepsilon^2$~\cite{Haah2016,ODonnell2016-1,ODonnell2016-2}. More recently, Ref.~\cite{PelecanosSpileckiTangWright2025} gave a particularly simple high-confidence procedure; its sample complexity of this procedure is summarized in the following lemma, which will be useful in the rest of our work.

\begin{lemma}[Mixed-state tomography]
\label{lemma:global_mixed_state_tomography_trace_distance}
For all $\varepsilon,\delta\in(0,1)$, there exists a tomography procedure which, given
\[
    N =
    O\left(
        \frac{D^2+\log(1/\delta)}{\varepsilon^2}
    \right)
\]
copies of an arbitrary unknown quantum state $\sigma\in\DD(\mathbb{C}^D)$, outputs an estimator $\widehat{\sigma}\in\DD(\mathbb{C}^D)$ such that
\[
    \Pr\left[
        \frac12\left\|\widehat{\sigma}-\sigma\right\|_1
        \leq \varepsilon
    \right]
    \geq 1-\delta .
\]
\end{lemma}

In order to obtain the proof of Theorem 1 of the main text, let us proceed by introducing the \emph{classical teleportation protocol}. Given a bipartite state $\rho$ shared between Alice and Bob, Alice can teleport a noisy version of her part to Bob with an LOCC procedure called classical teleportation. There are two ways to perform it:
\begin{itemize}
    \item \emph{First way}: Alice and Bob produce the separable isotropic state having maximal fidelity with the maximally entangled state, i.e., $W_d\coloneqq\frac{1}{d+1}\Phi_d+(1-\frac{1}{d+1})\frac{\id}{d^2}$~\cite{Horodecki1999}. 
    %\footnote{In order to show that the isotropic state $p\Phi_d+(1-p)\frac{\id}{d^2}$ with $p> \frac{1}{d+1}$ is entangled, it suffices to apply the PPT criterion~\cite{Horodecki1999}. Conversely, in order to show that the isotropic state $p\Phi_d+(1-p)\frac{\id}{d^2}$ with $0\le p\le \frac{1}{d+1}$ is separable, one can notice that it can be obtained by applying twirling to a suitable pure product state~\cite{Horodecki1999}, i.e., one can notice that for all $p\in[0,\frac{1}{d+1}]$ there exists a pure product state $\ketbra{\phi}\otimes\ketbra{\psi}$ such that $p\Phi_d+(1-p)\frac{\id}{d^2}=\int \dd U\, (U\otimes U^*)\, \ketbra{\phi}\otimes\ketbra{\psi} \, (U\otimes U^*)^\dag$. Hence, such an isotropic state can be written as a convex combination of product states, and is therefore separable.}. 
    This is allowed since the production of any separable state can be performed by means of a suitable LOCC. Alice and Bob perform the same steps of the quantum teleportation protocol~\cite{teleportation}, which aims at teleporting Alice's part of $\rho$ to Bob's laboratory, with the resource state to be consumed being $W_d$ instead of the maximally entangled state $\Phi_d$ as in the standard teleportation protocol. Using $W_d$ as the resource state for the quantum teleportation protocol is equivalent, at the level of the averaged output, to performing the quantum teleportation protocol with the resource state being: (i) $\Phi_d$ with probability $\frac{1}{d+1}$; (ii) $\frac{\id}{d^2}$ with probability $(1-\frac{1}{d+1})$. After such a protocol, in event (i), Bob gets exactly the state $\rho$, while in event (ii), Bob gets the state $\frac{\id_A}{d}\otimes \Tr_A\rho$. Hence, after such a protocol, Bob has the state $\frac{1}{d+1}\rho+(1-\frac{1}{d+1})\frac{\id_A}{d}\otimes \Tr_A\rho$.

    \item \emph{Second way}: Alice measures her part of the state $\rho_{AB}$ with respect to the POVM 
   \[ 
    \{ d\,\dd\mu(\psi)\ketbra{\psi}_A\, :\, \ket{\psi}_A\in\HH_A\},\] where $\dd\mu(\psi)$ denotes the Haar measure. Note that this is a POVM since each POVM element is positive semidefinite and since $d\int\dd\mu(\psi) \ketbra{\psi}_A =\id_A$. Alice obtains the outcome $\psi$ with probability $d\,\dd\mu(\psi)\Tr_{A}\left[\Tr_B \rho_{AB}\, \ketbra{\psi}_A\right]$. Then, Alice communicates the outcome $\psi$ to Bob, and he produces the state $\ketbra{\psi}_A$ in his laboratory. After such a procedure, Bob forgets the outcome $\psi$, and hence he has the state 
    \bb
    d\int \dd\mu(\psi)\bra{\psi}_A\rho_{AB}\, \ket{\psi}_A\ketbra{\psi}_A&= d\int \dd\mu(\psi)\bra{\psi}_{A'}\rho_{A'B}\, \ket{\psi}_{A'}\ketbra{\psi}_A\\&= d\Tr_{A'}\left[\int \dd\mu(\psi)\ketbra{\psi}_{A'}\otimes\ketbra{\psi}_A\, \rho_{A'B}\right]\\&=\frac{1}{d+1}\left[ \rho_{AB}+\id_{A}\otimes\Tr_A\rho_{AB} \right]\\&=\frac{1}{d+1}\rho_{AB}+\left(1-\frac{1}{d+1}\right)\frac{\id_A}{d}\otimes \Tr_A\rho_{AB}\,,
    \ee
    where we used that~\cite{Mele2024introductiontohaar}
    \bb
        \int \dd\mu(\psi)\ketbra{\psi}_{A'}\otimes\ketbra{\psi}_A=\binom{d+1}{2}^{-1}\left(\frac{\id_{A'A}+F_{A'A}}{2}\right)\,,
    \ee
    with $F_{A'A}$ being the flip operator defined by $F_{A'A}=\sum_{i,j=1}^d\ketbraa{j}{i}_{A'}\otimes\ketbraa{i}{j}_A$.
    
    Here we have presented the argument with a Haar-distributed POVM on Alice's side, whose outcome formally requires infinitely many bits to be communicated to Bob. However, the derivation only requires that Alice's POVM be formed from a finite (possibly weighted) projective two-design, so the same channel can in fact be implemented with finitely many bits of Alice-to-Bob communication.
\end{itemize}
At the end of such a classical teleportation procedure, the state in Bob's laboratory becomes 
\bb
    \mathcal{N}(\rho)\coloneqq\frac{1}{d+1}\rho+\left(1-\frac{1}{d+1}\right)\frac{\id_A}{d}\otimes \Tr_A\rho\,,
\ee
where we have introduced the quantum channel $\mathcal{N}$ defined as
\bb\label{definition_Phi}
    \mathcal{N}(\Theta)\coloneqq\frac{1}{d+1}\Theta+\left(1-\frac{1}{d+1}\right)\frac{\id_A}{d}\otimes\Tr_A[\Theta] \; 
\ee
for any linear operator $\Theta$ acting on $\HH_A\otimes\HH_B$. The quantum channel $\mathcal{N}$ is invertible if viewed as linear map, and its inverse is
\begin{equation}
    \mathcal{N}^{-1}(\Theta)
    =
    (d+1)\Theta-\id_{A}\otimes\Tr_A[\Theta]\,.
\label{eq:classical_teleportation_inverse}
\end{equation}
We would like to remark that the above linear map is not a quantum channel, but it is useful in classical post-processing. \\
With these results about global quantum state tomography and classical teleportation, we are now ready to state our LOCC tomography protocol for a bipartite state. The protocol is detailed in Table~\ref{table_LOCC_tomography}, and its correctness is proved in Theorem~\ref{thm:LOCC_tomography_qudit_qudit}.
\begin{table}[h!]
  \caption{LOCC tomography protocol for an unknown qudit--qudit state}
  \label{table_LOCC_tomography}
  \begin{mdframed}[linewidth=2pt, roundcorner=10pt, backgroundcolor=white!10, innerbottommargin=10pt, innertopmargin=10pt]
    \textbf{Input:}
    \begin{itemize}
        \item Accuracy parameter $\varepsilon\in(0,1)$;
        \item Failure probability $\delta\in(0,1)$;
        \item $N =
            O\left(
                \frac{d^6+d^2\log(1/\delta)}{\varepsilon^2}
            \right)$ copies of an unknown bipartite state
        $\rho_{AB}\in\DD(\mathbb{C}^{d}\otimes\mathbb{C}^{d})$ shared by
        Alice and Bob.
    \end{itemize}

    \textbf{Output:} A classical description of an estimator
    $\widehat{\rho}\in\DD(\mathbb{C}^{d}\otimes\mathbb{C}^{d})$ such that
    $\frac12\|\widehat{\rho}-\rho_{AB}\|_1\leq\varepsilon$ with probability at
    least $1-\delta$.

    \begin{algorithmic}[1]
    \State Set $\alpha\coloneqq\frac{\varepsilon}{2(2d+1)}$.
    \For{$j\leftarrow 1$ \textbf{to} $N$}
        \State Alice and Bob apply the classical teleportation procedure to the
        $j$-th copy of $\rho_{AB}$, so that Bob obtains in his laboratory a copy of the state $\mathcal{N}(\rho_{AB})$, where $\mathcal{N}$ is defined in Eq.~\eqref{definition_Phi}.
    \EndFor
    \State At this stage, Bob holds $N$ copies of $\mathcal{N}(\rho_{AB})$.
    \State Bob performs global mixed-state tomography on
    $\mathcal{N}(\rho_{AB})^{\otimes N}$ with target trace-distance accuracy
    $\alpha$ and failure probability $\delta$. Let $\widehat{\sigma}$ be the
    resulting estimator.
    \State Bob computes the raw inverse estimate explicitly as $\widetilde{\rho}
        \coloneqq
        \mathcal{N}^{-1}(\widehat{\sigma})
        =
        (d+1)\widehat{\sigma}
        -
        \id_{A}\otimes\Tr_A[\widehat{\sigma}]$.
    \State Diagonalize $\widetilde{\rho}=\sum_i\lambda_i\ketbra{i}$ and define $\widetilde{\rho}_{+}
        \coloneqq
        \sum_i \max\{\lambda_i,0\}\ketbra{i}$.
    \State Set $\widehat{\rho}
        \coloneqq
        \frac{\widetilde{\rho}_{+}}{\Tr \widetilde{\rho}_{+}}$.
    \State \Return $\widehat{\rho}$.
    \end{algorithmic}
  \end{mdframed}
\end{table}

\subsection{Proof of Theorem 1 of the main text}

Here, we give the formal statement and the formal proof of Theorem 1 in the main text.

\begin{thm}[Complete quantum tomography under LOCC operations]
\label{thm:LOCC_tomography_qudit_qudit}
For all $\varepsilon,\delta\in(0,1)$, there exists an LOCC tomography protocol which, given
\[
    N
    =
    O\left(
        \frac{d^6+d^2\log(1/\delta)}{\varepsilon^2}
    \right)
\]
copies of an unknown bipartite state $\rho_{AB}\in\DD(\mathbb{C}^{d}\otimes\mathbb{C}^{d})$, outputs an estimator
$\widehat{\rho}\in\DD(\mathbb{C}^{d}\otimes\mathbb{C}^{d})$ such that
\[
    \Pr\left[
        \frac12\left\|\widehat{\rho}-\rho_{AB}\right\|_1\leq\varepsilon
    \right]
    \geq 1-\delta .
\]
In particular, for constant failure probability $\delta$, the sample complexity is
$O(d^6/\varepsilon^2)$.
\end{thm}
\begin{proof}
We use the protocol described in Table~\ref{table_LOCC_tomography}. After the
classical teleportation step, Bob holds $N$ copies of
$\sigma_{AB}\coloneqq\mathcal{N}(\rho_{AB})$. This is a mixed state
on a Hilbert space of total dimension $D=d^2$. By
Lemma~\ref{lemma:global_mixed_state_tomography_trace_distance}, if
$N=O((D^2+\log(1/\delta))/\alpha^2)$, then Bob can construct an estimator
$\widehat{\sigma}$ such that, with probability at least $1-\delta$,
\begin{equation}
    \frac12
    \left\|
        \widehat{\sigma}-\mathcal{N}(\rho_{AB})
    \right\|_1
    \leq
    \alpha .
\label{eq:tomography_error_after_classical_teleportation}
\end{equation}
Let $X\coloneqq\widehat{\sigma}-\mathcal{N}(\rho_{AB})$, and define the raw
inverse estimate $\widetilde{\rho}\coloneqq\mathcal{N}^{-1}(\widehat{\sigma})$.
From Eq.~\eqref{eq:classical_teleportation_inverse},
\[
\widetilde{\rho}-\rho_{AB}
=
\mathcal{N}^{-1}(X)
=
(d+1)X-\id_{A}\otimes\Tr_A[X]\,.
\]
Since the trace norm is contractive under partial trace, we have
\begin{align}
    \left\|\mathcal{N}^{-1}(X)\right\|_1
    &\leq
    (d+1)\|X\|_1
    +
    \left\|
        \id_{A}\otimes\Tr_A[X]
    \right\|_1
    \nonumber\\
    &=
    (d+1)\|X\|_1
    +
    d\left\|\Tr_A[X]\right\|_1
    \nonumber\\
    &\leq
    (2d+1)\|X\|_1 .
\label{eq:inverse_channel_trace_norm_bound}
\end{align}

Thus, on the event in
Eq.~\eqref{eq:tomography_error_after_classical_teleportation}, the raw inverse
estimate $\widetilde{\rho}$ satisfies
\bb
    \frac12
    \left\|
        \widetilde{\rho}-\rho_{AB}
    \right\|_1
    \leq
    (2d+1)\alpha .
\ee

The operator $\widetilde{\rho}$ need not be positive semidefinite, because
$\mathcal{N}^{-1}$ is not a quantum channel. We correct this explicitly by
discarding its negative part and renormalizing. Namely, let
$\widetilde{\rho}=\widetilde{\rho}_{+}-\widetilde{\rho}_{-}$ be the Jordan
decomposition of $\widetilde{\rho}$, and define
\bb
    \widehat{\rho}
    \coloneqq
    \frac{\widetilde{\rho}_{+}}{\Tr\widetilde{\rho}_{+}} .
\ee
Since $\widehat{\sigma}$ has unit trace and $\mathcal{N}^{-1}$ is trace
preserving, we have $\Tr\widetilde{\rho}=1$. 
%Thus, w
Writing $s\coloneqq\Tr\widetilde{\rho}_{-}$ gives
$\Tr\widetilde{\rho}_{+}=1+s$. Note also that $\widehat{\rho}$ is well-defined and is a
density operator.

Let us now bound the error introduced by this truncation and renormalization.
Let $P_-$ be the projector onto the support of $\widetilde{\rho}_{-}$. Since
$\rho_{AB}\geq0$, we have $\Tr[P_-\rho_{AB}]\geq0$, while
$\Tr[P_-\widetilde{\rho}]=-\Tr\widetilde{\rho}_{-}=-s$. Therefore,
\bb
    s
    \leq
    \Tr[P_-(\rho_{AB}-\widetilde{\rho})]
    \leq
    \frac12\left\|\rho_{AB}-\widetilde{\rho}\right\|_1 ,
\ee
where in the last step we used the variational characterization
$\frac12\|H\|_1=\sup_{0\le E\le\id}\Tr[HE]$ for Hermitian traceless
operators $H$, with $H=\rho_{AB}-\widetilde{\rho}$ and $E=P_-$. Moreover,
\bb
    \left\|
        \widehat{\rho}-\widetilde{\rho}_{+}
    \right\|_1
    =
    \left\|
        \frac{\widetilde{\rho}_{+}}{1+s}
        -
        \widetilde{\rho}_{+}
    \right\|_1
    =
    s,
    \qquad
    \left\|
        \widetilde{\rho}_{+}-\widetilde{\rho}
    \right\|_1
    =
    s.
\ee
Hence,
\bb
    \left\|
        \widehat{\rho}-\widetilde{\rho}
    \right\|_1
    \leq
    2s
    \leq
    \left\|
        \widetilde{\rho}-\rho_{AB}
    \right\|_1 .
\ee
By the triangle inequality,
\bb
    \left\|
        \widehat{\rho}-\rho_{AB}
    \right\|_1
    \leq
    2\left\|
        \widetilde{\rho}-\rho_{AB}
    \right\|_1 .
\ee
Therefore,
\bb
    \frac12
    \left\|
        \widehat{\rho}-\rho_{AB}
    \right\|_1
    \leq
    \left\|
        \widetilde{\rho}-\rho_{AB}
    \right\|_1
    \leq
    2(2d+1)\alpha .
\ee
Choosing $\alpha=\varepsilon/(2(2d+1))$ gives
$\frac12\|\widehat{\rho}-\rho_{AB}\|_1\leq\varepsilon$ with probability at
least $1-\delta$.

It remains to substitute this value of $\alpha$ into the sample complexity.
Since $D=d^2$,
\begin{align}
    N
    &=
    O\left(
        \frac{D^2+\log(1/\delta)}{\alpha^2}
    \right) \nonumber \\
   & =
    O\left(
        \frac{d^4+\log(1/\delta)}
        {\varepsilon^2/(4(2d+1)^2)}
    \right)
    \nonumber\\
    &=
    O\left(
        \frac{d^6+d^2\log(1/\delta)}{\varepsilon^2}
    \right).
\end{align}
For constant failure probability, this is $O(d^6/\varepsilon^2)$.
\end{proof}

The same argument also applies when the local dimensions are different. Writing
\[
    d\coloneqq\min\{|A|,|B|\},\qquad D\coloneqq|A||B|,
\]
and classically teleporting the smaller subsystem gives sample complexity
\bb
    N=O\!\left(\frac{d^2D^2+d^2\log(1/\delta)}{\varepsilon^2}\right) .
\label{eq:unequal_dimension_tomography}
\ee
Indeed, the global tomography now acts on a \(D\)-dimensional state, while the inverse-channel estimate still has the factor \(2d+1\).

\section{Uniform hashing for an isotropic shell}

In this Section, we prove a hashing protocol that allows Alice and Bob to distill ebits uniformly on a local subset of states. This result is essential in the local robustness Lemma and in the proof of Theorem~\ref{thm:pointwise-universal-distillation}. The proof is the standard sequential hashing proof of entanglement purification, specialized to isotropic states and written in a form that is uniform over a compact interval of fidelities; see Refs.~\cite{Bennett-error-correction,Vollbrecht2003,devetak2005}. Fix an integer \(\ell\ge1\). On a \(2^\ell\times2^\ell\) bipartite system, the isotropic twirling channel is LOCC, since Alice and Bob sample the unitary using shared randomness and apply it locally. The channel preserves the overlap with \(\Phi_2^{\otimes\ell}\). Indeed, if \(D=2^\ell\) and \(\Psi_D=\Phi_2^{\otimes\ell}\), then
\[
    (U\otimes U^*)\ket{\Psi_D}=\ket{\Psi_D},
\]
and hence, for every operator \(X\),
\[
    \Tr[\Psi_D\,\TT(X)]
    =
    \int \dd U\,\Tr[(U\otimes U^*)^\dagger\Psi_D(U\otimes U^*)X]
    =
    \Tr[\Psi_DX].
\]
By the symmetry of the twirl, \(\TT(X)\) is a linear combination of \(\Psi_D\) and \(\id-\Psi_D\). Therefore, if $X$ is a state and \(q=\Tr[\Psi_DX]\), the twirled state is exactly the isotropic state below. Let
\[
    \tau_{2^\ell}
    =
    \frac{\id-\Phi_2^{\otimes\ell}}{4^\ell-1} .
\]
For \(q\in[0,1]\), define
\begin{equation}
    \omega_q
    \coloneqq
    q\,\Phi_2^{\otimes\ell}
    +(1-q)\tau_{2^\ell} .
\label{eq:isotropic_omega}
\end{equation}
Let
\[
    h_2(q)=-q\log q-(1-q)\log(1-q)
\]
be the binary entropy, with the convention \(0\log0=0\).

\begin{lemma}[Uniform hashing for an isotropic shell]
\label{lem:uniform_hashing}
Fix an integer \(\ell\ge1\), and let \(q_0\in[4^{-\ell},1]\). For every rate
\bb
    0<R_h<I(A\rangle B)_{\omega_{q_0}}
    =\ell-h_2(q_0)-(1-q_0)\log(4^\ell-1),
\label{eq:uniform_hashing_rate}
\ee
there exists a sequence of LOCC maps \(\mathsf H_N\) such that
\[
    \lim_{N\to\infty}
    \sup_{q\in[q_0,1]}
    \half
    \norm{\mathsf H_N(\omega_q^{\otimes N})-
    \Phi_2^{\otimes\ceil{R_hN}}}_1
    =0 .
\]

We will also need the following finite-block estimate. Let \(N\ge1\) and \(t>0\) satisfy
\[
    r\coloneqq1-q_0+t\le1-4^{-\ell},
\]
and set
\bb
    s_N\coloneqq
    \left\lceil N\bigl[h_2(r)+r\log(4^\ell-1)+t\bigr]\right\rceil .
\label{eq:syndrome_length}
\ee
Whenever \(m_N\coloneqq N\ell-s_N>0\), there is an LOCC map \(\widetilde{\mathsf H}_N\), independent of \(q\), such that
\bb
    \sup_{q\in[q_0,1]}
    \half\norm{
        \widetilde{\mathsf H}_N(\omega_q^{\otimes N})
        -\Phi_2^{\otimes m_N}
    }_1
    \le \exp\left[-2Nt^2\right]+2^{-Nt} .
\label{eq:uniform_hashing_finite_block}
\ee
\end{lemma}

\begin{proof}
We give the proof explicitly because the sequential update of the Bell labels is essential. After the twirling operation, the procedure reduces to the classical problem of learning enough about a Bell-error string to correct the Bell pairs that have not been consumed by the parity measurements.

\medskip
\noindent\textbf{Step 1: the Bell-error variable and a uniform candidate list.}
Let \(D\coloneqq2^\ell\). Choose a tensor-product Pauli Bell basis \(\{\Xi_z:z\in\mathcal Z\}\) of the \(D\times D\) system, where \(\mathcal Z\simeq\mathbb F_2^{2\ell}\), $\mathbb{F}_2$ is the Galois field with two elements, \(|\mathcal Z|=D^2\), and \(\Xi_0=\Phi_2^{\otimes\ell}\). Then
\begin{equation}
    \omega_q
    =
    \sum_{z\in\mathcal Z}p_q(z)\Xi_z,
\label{eq:omega_bell_diagonal}
\end{equation}
where
\begin{equation}
    p_q(0)=q,
    \qquad
    p_q(z)=\frac{1-q}{D^2-1}
    \quad(z\ne0).
\label{eq:pq_distribution}
\end{equation}
For \(N\) copies, write
\[
    Z^N=(Z_1,\ldots,Z_N)\in\mathcal Z^N
\]
for the corresponding Bell-error string. Its one-copy entropy is
\begin{equation}
    H(p_q)
    =
    h_2(q)+(1-q)\log(D^2-1).
\label{eq:pq_entropy}
\end{equation}
Moreover,
\begin{equation}
    \frac{\dd}{\dd q}H(p_q)
    =
    \log\frac{1-q}{q(D^2-1)} .
\label{eq:entropy_derivative}
\end{equation}
Since \(q_0\ge D^{-2}\), this derivative is nonpositive throughout \([q_0,1)\), and therefore, by continuity at $q=1$,
\begin{equation}
    H(p_q)\le H(p_{q_0})
    \qquad(q\in[q_0,1]).
\label{eq:uniform_entropy_bound}
\end{equation}
The reduced states of \(\omega_q\) are maximally mixed, so \(I(A\rangle B)_{\omega_q}=\ell-H(p_q)\). Thus the right-hand side of \eqref{eq:uniform_hashing_rate} is precisely the smallest coherent information on the shell.

We first prove the finite-block estimate. Fix \(N,t,r\) as in the statement, with \(s_N<N\ell\). For a deterministic string \(z^N\), let \(w(z^N)\) be the number of its nonzero block labels, and define the candidate list
\bb
    \mathcal A_N
    \coloneqq
    \left\{z^N\in\mathcal Z^N:w(z^N)\le Nr\right\} .
\label{eq:AN_definition}
\ee
This list is independent of \(q\). Since \(r\le1-D^{-2}\), every string in the list has probability at least \(2^{-N[h_2(r)+r\log(D^2-1)]}\) under \(p_{1-r}^{\otimes N}\). Indeed, the probability of a string of weight \(j\le Nr\) is
\[
    (1-r)^{N-j}\left(\frac{r}{D^2-1}\right)^j
    \ge
    (1-r)^{N(1-r)}\left(\frac{r}{D^2-1}\right)^{Nr} .
\]
Summing these probabilities gives
\bb
    |\mathcal A_N|
    =\sum_{j=0}^{\floor{Nr}}\binom Nj(D^2-1)^j
    \le 2^{N[h_2(r)+r\log(D^2-1)]} .
\label{eq:AN_size}
\ee
Also, under \(p_q^{\otimes N}\), the weight \(w(Z^N)\) is a sum of \(N\) independent Bernoulli variables with mean \(1-q\le1-q_0=r-t\). Hoeffding's inequality~\cite{Hoeffding1963} therefore gives
\bb
    \sup_{q\in[q_0,1]}
    \Pr_{p_q}[Z^N\notin\mathcal A_N]
    \le \exp[-2Nt^2] .
\label{eq:uniform_typicality}
\ee
Neither estimate has a prefactor that depends on the alphabet; in particular, they remain valid when \(\ell\) varies with \(N\).

\medskip
\noindent\textbf{Step 2: the one-round parity-extraction primitive.}
Identify the \(N\ell\) Bell pairs with a Bell label
\[
    x\in\mathbb F_2^{2N\ell}
\]
obtained by concatenating the block labels in \(z^N\). More generally, suppose that \(m\ge1\) Bell pairs remain and that their current Bell label is \(x\in\mathbb F_2^{2m}\). For every nonzero vector \(h\in\mathbb F_2^{2m}\), there is a bilateral Clifford operation followed by local measurements and classical communication which does the following:
\begin{enumerate}
    \item it reveals the single parity bit \(h\cdot x\);
    \item it consumes one Bell pair;
    \item it leaves the other \(m-1\) pairs Bell diagonal, with a label
    \[
        f_h(x)\in\mathbb F_2^{2(m-1)}
    \]
    that is a known linear function of \(x\).
\end{enumerate}

Here is an elementary way to obtain this primitive. Up to phases, write a Pauli string as
\[
    P_x=\bigotimes_{j=1}^m X^{u_j}Z^{v_j},
    \qquad x=(u_1,v_1,\ldots,u_m,v_m),
\]
and label the Bell basis by \(\ket{\Xi_x}=(\id\otimes P_x)\ket{\Phi_2}^{\otimes m}\). For every nonzero \(h\), there is a nonidentity Pauli string \(Q_h\) whose commutation relation with \(P_x\) is
\[
    Q_hP_x=(-1)^{h\cdot x}P_xQ_h .
\]
Explicitly, if \(h=(h_{u,1},h_{v,1},\ldots,h_{u,m},h_{v,m})\), take \(Q_h=\bigotimes_j X^{h_{v,j}}Z^{h_{u,j}}\), up to a phase.

The Clifford group acts transitively on nonidentity Pauli strings, up to phases. To see this, first use single-qubit Clifford gates to turn every nonidentity factor of \(Q_h\) into \(X\). If \(Q_h\) has two\(X\) factors, a CNOT with those factors as control and target removes the \(X\) on the target. Repeating this procedure leaves just one \(X\), which can be moved to the designated target pair using SWAPs and turned into \(Z\) by a Hadamard gate. Thus, we can choose a Clifford \(C\) with \(CQ_hC^\dagger=Z\) on the target qubit, up to a phase.

Alice and Bob apply \(C^*\otimes C\). This replaces \(P_x\) by \(CP_xC^\dagger\), whose amplitude bit on the target pair is \(h\cdot x\), by the commutation relation above. Measuring that pair locally in the computational basis and comparing the outcomes therefore reveals the required parity. Clifford conjugation preserves Pauli multiplication up to phases, while multiplication of Pauli strings adds their binary labels. Hence, the transformed label is a known invertible linear function of \(x\). Discarding the measured pair simply deletes its two label coordinates, leaving the known linear function \(f_h(x)\) on the other pairs. We fix one such Clifford for every nonzero \(h\), so that \(f_h\) is fixed as well. This protocol is the standard one-round primitive used in the hashing protocol~\cite{Bennett-error-correction,Vollbrecht2003}.

This primitive has the following property. Let \(x\ne y\) and choose \(h\) uniformly from \(\mathbb F_2^{2m}\setminus\{0\}\). The two parity outcomes coincide precisely when \(h\cdot(x-y)=0\). In this particular instance, we say that the two strings \textit{collide}. Fixing $z\coloneqq x-y\neq 0$, the string $h$ can take $2^{2m}-1$ different values (the all-$0$ string is excluded), and exactly $2^{2m-1}-1$ of those, i.e., those that have an even number of $1$'s in the positions where also $z$ has $1$'s, lead to $h\cdot z=0$. Hence
\begin{equation}
    \Pr_h[h\cdot x=h\cdot y]
    =
    \frac{2^{2m-1}-1}{2^{2m}-1}
    <\frac12 .
\label{eq:one_round_collision}
\end{equation}
If the outcomes coincide and \(f_h(x)=f_h(y)\), the two hypotheses have merged into the same residual Bell label and no longer need to be distinguished. If the outcomes coincide but the residual labels remain different, another round is required.

\medskip
\noindent\textbf{Step 3: the sequential random hashing protocol.}
Starting from \(M_N\coloneqq N\ell\) Bell pairs, Alice and Bob perform \(s_N\) rounds. At round \(k\), when \(M_N-k\) pairs remain, they use shared randomness to choose
\[
    h_k\in\mathbb F_2^{2(M_N-k)}\setminus\{0\}
\]
uniformly and independently of the previous choices, and then apply the one-round primitive associated with \(h_k\). Let \(b_k\) be the measured parity bit. For an initial candidate \(z^N\), denote by
\[
    x_0(z^N),x_1(z^N),\ldots,x_{s_N}(z^N)
\]
its successively updated Bell labels and by
\[
    b^{s_N}(z^N)=(b_0(z^N),\ldots,b_{s_N-1}(z^N))
\]
the parity bits predicted for this candidate, using the actual random choices \(h_0,\ldots,h_{s_N-1}\) made in the protocol. These bits, which we call its \emph{simulated transcript}, can be calculated classically. The update is sequential:
\[
    x_{k+1}(z^N)=f_{h_k}(x_k(z^N)),
    \qquad
    b_k(z^N)=h_k\cdot x_k(z^N).
\]
Thus the measured parities are parities of the \emph{current} Bell labels.
Fix two distinct initial candidates \(z^N,z'^N\). For
\(k=0,\ldots,s_N\), write
\[
    b^k(z^N)
    \coloneqq
    \bigl(b_0(z^N),\ldots,b_{k-1}(z^N)\bigr),
\]
with \(b^0(z^N)\) denoting the empty transcript, and define
\begin{equation}
    E_k
    \coloneqq
    \left\{
        b^k(z^N)=b^k(z'^N),
        \quad
        x_k(z^N)\ne x_k(z'^N)
    \right\}.
\label{eq:Ek_definition}
\end{equation}
Since the initial candidates are distinct and the map from an initial
Bell-error string to its concatenated Bell label is injective,
\[
    x_0(z^N)\ne x_0(z'^N),
\]
and hence \(E_0\) occurs with probability one.

We first note that
\begin{equation}
    E_{k+1}\subseteq E_k .
\label{eq:Ek_nested}
\end{equation}
Indeed, agreement of the first \(k+1\) transcript bits implies agreement
of the first \(k\) transcript bits. Moreover, if
\(x_k(z^N)=x_k(z'^N)\), the two candidates are subjected to the same
deterministic update \(f_{h_k}\), and therefore
\[
    x_{k+1}(z^N)
    =
    f_{h_k}\!\left(x_k(z^N)\right)
    =
    f_{h_k}\!\left(x_k(z'^N)\right)
    =
    x_{k+1}(z'^N).
\]
Thus two residual labels that have merged can never become different
again.

For any fixed history \(h_0,\ldots,h_{k-1}\) for which \(E_k\) occurs, the two current labels are fixed and distinct. The next choice \(h_k\) is independent of that history and uniformly distributed over \(\mathbb F_2^{2(M_N-k)}\setminus\{0\}\). By Eq.~\eqref{eq:one_round_collision}, the probability that their next parity bits agree is
\[
    \frac{2^{2(M_N-k)-1}-1}{2^{2(M_N-k)}-1}<\frac12 .
\]
Requiring the updated residual labels to remain different can only reduce this probability. Averaging over the histories for which \(E_k\) occurs, and using \(E_{k+1}\subseteq E_k\), gives
\bb
    \Pr[E_{k+1}]
    \le
    \frac{2^{2(M_N-k)-1}-1}{2^{2(M_N-k)}-1}\Pr[E_k]
    \le\frac12\Pr[E_k] .
\label{eq:Ek_conditional_recursion}
\ee
If \(E_k\) has probability zero, both sides vanish.
Iterating this inequality from \(k=0\) to \(k=s_N-1\), and using
\(\Pr[E_0]=1\), yields
\begin{equation}
    \Pr[E_{s_N}]
    \le
    2^{-s_N}.
\label{eq:Ek_iteration}
\end{equation}
More precisely, the preceding argument gives the slightly stronger bound
\[
    \Pr[E_{s_N}]
    \le
    \prod_{k=0}^{s_N-1}
    \frac{2^{2(M_N-k)-1}-1}
         {2^{2(M_N-k)}-1}
    <
    2^{-s_N}.
\]
Finally, by the definition of \(E_{s_N}\),
\[
    E_{s_N}
    =
    \left\{
      b^{s_N}(z^N)=b^{s_N}(z'^N),
      \quad
      x_{s_N}(z^N)\ne x_{s_N}(z'^N)
    \right\}.
\]
Consequently,
\begin{equation}
    \Pr\!\left[
      b^{s_N}(z^N)=b^{s_N}(z'^N),\ 
      x_{s_N}(z^N)\ne x_{s_N}(z'^N)
    \right]
    \le2^{-s_N}.
\label{eq:sequential_collision_bound}
\end{equation}
The probability is over the shared random choices
\(h_0,\ldots,h_{s_N-1}\). No independence between the events
\(E_0,E_1,\ldots,E_{s_N}\) is being assumed. 
% the conclusion follows from their nesting and from the uniform conditional one-round bound.

\medskip
\noindent\textbf{Step 4: decoding the residual Bell label.}
After the \(s_N\) rounds, Alice and Bob know the random choices \(h_0,\ldots,h_{s_N-1}\) and the observed transcript \(b^{s_N}\). They simulate the sequential update for every candidate in \(\mathcal A_N\) and retain those candidates whose simulated transcript equals the observed one. If all retained candidates have the same residual label \(\widetilde x\), Bob applies the corresponding Pauli correction to the remaining
\begin{equation}
    m_N\coloneqq N\ell-s_N
\label{eq:kept_bell_pairs}
\end{equation}
Bell pairs. If there is no retained candidate, or if the retained candidates have more than one residual label, the protocol applies an arbitrary correction.

For a fixed true string \(z^N\in\mathcal A_N\), decoding can fail only if there exists \(z'^N\in\mathcal A_N\), \(z'^N\ne z^N\), which has the same transcript but a different final residual label. By the union bound and \eqref{eq:sequential_collision_bound},
\begin{equation}
    \Pr[\text{decoding failure}\mid Z^N=z^N]
    \le
    |\mathcal A_N|2^{-s_N}.
\label{eq:decoding_given_typical_string}
\end{equation}
Therefore, for every \(q\in[q_0,1]\), the total decoding-error probability \(p_N(q)\) satisfies
\begin{align}
    p_N(q)
    &\le
    \Pr_{p_q}[Z^N\notin\mathcal A_N]
    +|\mathcal A_N|2^{-s_N}
    \nonumber\\
    &\le
    \exp[-2Nt^2]+2^{-Nt}.
\label{eq:decoding_error_uniform_bound}
\end{align}
For fixed \(t>0\), the right-hand side is independent of \(q\) and tends to zero. Hence
\begin{equation}
    \lim_{N\to\infty}
    \sup_{q\in[q_0,1]}p_N(q)=0.
\label{eq:uniform_decoding_error}
\end{equation}

\medskip
\noindent\textbf{Step 5: from decoding success to Bell pairs.}
Denote by \(\widetilde{\mathsf H}_N\) the LOCC map consisting of the sequential parity measurements, decoding, and final Pauli correction, before any rate-adjusting discard. On every branch on which the residual Bell label is decoded correctly, this map produces exactly \(m_N\) perfect ebits. Thus, after averaging over the Bell-error string, the measurement outcomes, and the shared randomness, the output has the form
\begin{equation}
    (1-p_N(q))\Phi_2^{\otimes m_N}
    +p_N(q)\theta_{N,q}
\label{eq:output_success_failure_decomposition}
\end{equation}
for some state \(\theta_{N,q}\). Consequently,
\begin{equation}
    \half
    \norm{
        \widetilde{\mathsf H}_N(\omega_q^{\otimes N})
        -\Phi_2^{\otimes m_N}
    }_1
    \le p_N(q).
\label{eq:quantum_error_from_decoding_error}
\end{equation}
Together with \eqref{eq:decoding_error_uniform_bound}, this proves the finite-block estimate~\eqref{eq:uniform_hashing_finite_block}. For fixed \(t>0\), it also gives
\begin{equation}
    \lim_{N\to\infty}
    \sup_{q\in[q_0,1]}
    \half
    \norm{
        \widetilde{\mathsf H}_N(\omega_q^{\otimes N})
        -\Phi_2^{\otimes m_N}
    }_1
    =0.
\label{eq:uniform_hashing_to_mN}
\end{equation}

\medskip
\noindent\textbf{Step 6: rate check and discarding surplus Bell pairs.}
Let \(R_h\) satisfy \eqref{eq:uniform_hashing_rate}. By continuity of the binary entropy, we can fix \(t>0\) so small that \(r=1-q_0+t\le1-4^{-\ell}\) and
\bb
    R_h<\ell-h_2(r)-r\log(4^\ell-1)-t .
\label{eq:hashing_rate_margin}
\ee
From \eqref{eq:syndrome_length},
\bb
    m_N
    \ge N\bigl[\ell-h_2(r)-r\log(4^\ell-1)-t\bigr]-1 .
\label{eq:mN_lower_bound_first}
\ee
Hence \(m_N\ge\ceil{R_hN}\) for all sufficiently large \(N\). Let \(D_N\) be the LOCC channel that discards the surplus Bell pairs, and define \(\mathsf H_N\coloneqq D_N\circ\widetilde{\mathsf H}_N\). Since discarding cannot increase trace distance, Eq.~\eqref{eq:uniform_hashing_finite_block} yields
\[
    \lim_{N\to\infty}
    \sup_{q\in[q_0,1]}
    \half
    \norm{
        \mathsf H_N(\omega_q^{\otimes N})
        -\Phi_2^{\otimes\ceil{R_hN}}
    }_1
    =0.
\]
For the finitely many remaining blocklengths, let \(\mathsf H_N\) output a fixed product state in the required output space. This proves the lemma.
\end{proof}

\section{Local robustness and lower semicontinuity of the distillable entanglement}

We now prove the key technical result. Any rate strictly below \(E_d(\rho)\) is achievable not only at the single point \(\rho\), but uniformly on a whole trace-distance ball around \(\rho\), using one fixed sequence of LOCC maps.

\begin{lemma}[Local robustness]
\label{lem:local_universal_robustness}
Let \(\rho\in\DD(\mathbb C^d\otimes\mathbb C^d)\), and let \(0<R<E_d(\rho)\). Then there exist \(r>0\) and a sequence of LOCC maps \(\Lambda_n^{(\rho,R)}\) such that
\begin{equation}
    \lim_{n\to\infty}
    \sup_{\sigma\in B_r(\rho)}
    \half
    \norm{\Lambda_n^{(\rho,R)}(\sigma^{\otimes n})-
    \Phi_2^{\otimes\ceil{Rn}}}_1
    =0 .
\label{eq:local_robustness}
\end{equation}
\end{lemma}

\begin{proof}
Choose two intermediate rates
\[
    R<R_0<R_1<E_d(\rho).
\]
We first choose the accuracy parameter for the finite block. Pick \(\eta\in(0,1/4)\) so small that
\begin{equation}
    4\eta R_1 < \frac{R_1-R_0}{3} .
\label{eq:eta_small_first}
\end{equation}
This choice is made before choosing the block length.

Since \(R_1<E_d(\rho)\), the definition of \(E_d(\rho)\) gives, for arbitrarily large block lengths, LOCC maps that distill at rate \(R_1\) with error tending to zero. Choose one such block length \(k\), large enough for the two requirements below, and choose an LOCC map \(\Gamma\) with
\[
    \ell\coloneqq\ceil{R_1k},
    \qquad
    \Phi\coloneqq\Phi_2^{\otimes\ell},
\]
such that
\begin{equation}
    \Tr\!\left[\Phi\,\Gamma(\rho^{\otimes k})\right]
    \ge
    1-\eta .
\label{eq:block_good_at_center}
\end{equation}
%This is possible because vanishing trace-distance error implies fidelity tending to one, and for a pure target \(\Phi\), the squared fidelity equals the overlap \(\Tr[\Phi(\cdot)]\).

The block length \(k\) is also chosen large enough so that, with
\[
    q_0\coloneqq1-2\eta,
\]
one has
\begin{equation}
    \ell-\bigl[h_2(q_0)+(1-q_0)\log(4^\ell-1)\bigr]
    >
    R_0k .
\label{eq:hashing_margin}
\end{equation}
Let us justify that this second requirement can be met. Since \(1-q_0=2\eta\),
\[
    h_2(q_0)+(1-q_0)\log(4^\ell-1)
    \le
    h_2(2\eta)+4\eta\ell .
\]
Also \(\ell\ge R_1k\) and \(\ell\le R_1k+1\). Hence
\[
\begin{aligned}
\ell-&\bigl[h_2(q_0)+(1-q_0)\log(4^\ell-1)\bigr] 
\ge
R_1k-h_2(2\eta)-4\eta(R_1k+1).
\end{aligned}
\]
By \eqref{eq:eta_small_first}, the coefficient of \(k\) on the right-hand side is strictly larger than \(R_0\). Taking \(k\) large enough makes the constant terms negligible, proving \eqref{eq:hashing_margin}.

Now choose
\[
    0<r\le\frac{\eta}{k}.
\]
For every \(\sigma\in B_r(\rho)\), Lemma~\ref{lem:fixed_map_continuity} and \eqref{eq:block_good_at_center} give
\[
    \Tr\!\left[\Phi\,\Gamma(\sigma^{\otimes k})\right]
    \ge
    1-2\eta
    =q_0 .
\]
Alice and Bob then apply the isotropic twirling channel to the output of \(\Gamma\). Twirling is LOCC and preserves the overlap with \(\Phi\). Therefore each block is transformed into an isotropic state \(\omega_{q(\sigma)}\) of the form \eqref{eq:isotropic_omega}, with
\[
    q(\sigma)\in[q_0,1].
\]
The important point is that \(\TT\circ\Gamma\) is fixed. It depends only on \(\rho\) and on \(R\), not on the unknown \(\sigma\).

Now take \(n\) input copies of \(\sigma\). Split \(k\floor{n/k}\) of them into
\[
    N_n\coloneqq\floor{n/k}
\]
blocks of size \(k\), and discard the remaining fewer than \(k\) copies. Applying \(\TT\circ\Gamma\) to every block produces
\[
    \omega_{q(\sigma)}^{\otimes N_n},
    \qquad q(\sigma)\in[q_0,1].
\]
By \eqref{eq:hashing_margin} and Lemma~\ref{lem:uniform_hashing}, there is a sequence of LOCC hashing maps that distills from all states \(\omega_q\), \(q\in[q_0,1]\), uniformly at rate \(R_0k\) Bell pairs per block. Thus
\begin{equation}
    \lim_{n\to\infty}\sup_{\sigma\in B_r(\rho)}
    \half
    \norm{
    \mathsf H_{N_n}\!\left(\omega_{q(\sigma)}^{\otimes N_n}\right)
    -
    \Phi_2^{\otimes\ceil{R_0kN_n}}
    }_1 = 0 .
\label{eq:local_before_discard}
\end{equation}
Since
\[
    \lim_{n\to\infty}\frac{R_0kN_n}{n} = R_0 > R,
\]
for all sufficiently large \(n\) one has
\[
    \ceil{R_0kN_n}\ge\ceil{Rn} .
\]
Let \(D_n\) be the LOCC channel that discards the surplus Bell pairs, so that
\[
    D_n\!\left(\Phi_2^{\otimes\ceil{R_0kN_n}}\right)
    =
    \Phi_2^{\otimes\ceil{Rn}} .
\]
Define \(\Lambda_n^{(\rho,R)}\) to be the composition of block partitioning, \((\TT\circ\Gamma)^{\otimes N_n}\), hashing, and the final discard map \(D_n\). By data processing for the trace norm, applying \(D_n\) to \eqref{eq:local_before_discard} cannot increase the trace distance with the target state. Hence \eqref{eq:local_robustness} follows. For the finitely many remaining blocklengths, let $\Lambda_n^{(\rho,R)}$ output a fixed product state in the required output space.
\end{proof}

\begin{corollary}[Lower semicontinuity of the distillable entanglement]
\label{cor:Ed-lower-semicontinuity}
Let $\rho$ be a finite-dimensional bipartite state. Then, for every rate \(R<E_d(\rho)\), there exists \(r_R>0\) such that
\bb \label{eq:low_semicont}
    E_d(\sigma)>R
    \qquad\text{for all }\sigma\in B_{r_R}(\rho).
\ee
In particular, for every sequence of states $\{\sigma_n\}_{n\in\mathbb N}$ such that $\lim_{n\to\infty} \left\|\sigma_n - \rho\right\|_1 = 0$, it holds that
\bb
    \liminf_{n\to\infty} E_d(\sigma_n)\ge E_d(\rho).
\ee

\end{corollary}

\begin{proof}
We start by proving the first inequality. Tha case in which \(R<0\) follows immediately from the nonnegativity of \(E_d\), and any positive radius may be chosen. Assume now that \(0\le R<E_d(\rho)\). In particular, \(E_d(\rho)>0\), so we may choose a second rate \(R'\) with
\[
    R<R'<E_d(\rho).
\]
Lemma~\ref{lem:local_universal_robustness}, applied at the positive rate \(R'\), gives a radius \(r_R>0\) and a sequence of LOCC maps which distills at rate \(R'\) with vanishing error uniformly for all \(\sigma\in B_{r_R}(\rho)\). Hence \(R'\) is an achievable distillation rate for every such \(\sigma\), and therefore
\[
    E_d(\sigma)\ge R'>R
    \qquad\text{for all }\sigma\in B_{r_R}(\rho).
\]

Now let $\{\sigma_n\}_{n\in\mathbb N}$ be a sequence of states that converges to $\rho$ in trace distance. If \(E_d(\rho)=0\), then
\[
    \liminf_{n\to\infty}E_d(\sigma_n)\ge 0=E_d(\rho)
\]
by nonnegativity. If \(E_d(\rho)>0\), then for every \(0\le R<E_d(\rho)\), the already proved inequality \eqref{eq:low_semicont}  implies that \(E_d(\sigma_n)>R\) for all sufficiently large \(n\). Thus
\[
    \liminf_{n\to\infty}E_d(\sigma_n)\ge R.
\]
Letting \(R\uparrow E_d(\rho)\) gives
\[
    \liminf_{n\to\infty}E_d(\sigma_n)\ge E_d(\rho),
\]
thereby concluding the proof.
\end{proof}

\section{Proof of Theorem 3 of the main text}

We now prove the main statement of our work: there is one universal variable-rate protocol which, for each fixed unknown input state, achieves every rate strictly below the ordinary distillable entanglement of that state.

The delicate point of this proof is the number of copies used during the tomography step. The number of copies spent on tomography must remain sublinear in the total block length, but it also must be large enough to give the requested tomographic accuracy needed to use the local robustness Lemma. 

\begin{thm}[Universal distillation]
\label{thm:pointwise-universal-distillation}
There exists a single sequence of LOCC maps with fixed output spaces \(\mathcal U_n\), independent of the input state, whose output consists of a classical length register and a padded finite register of qubit pairs. For every state
\(\rho\in\DD(\mathbb C^d\otimes\mathbb C^d)\) and every rate
\(0<R<E_d(\rho)\), there is a deterministic LOCC readout map \(\mathcal D_{n,R}\), depending only on \(n\) and \(R\), such that
\[
    \lim_{n\to\infty}
    \half
    \norm{
        \mathcal D_{n,R}\!\circ\mathcal U_n(\rho^{\otimes n})
        -\Phi_2^{\otimes\ceil{Rn}}
    }_1
    =0.
\]
The map \(\mathcal D_{n,R}\) merely keeps the first \(\ceil{Rn}\) declared Bell-pair registers when enough are available, and otherwise outputs a fixed separable state. Thus the universal part \(\mathcal U_n\) is independent of both \(\rho\) and the target rate. Its error averaged over the declared output length tends to zero for every state; moreover, for every $0<R<E_d(\rho)$, the probability of declaring fewer than $\ceil{Rn}$ pairs tends to zero.
\end{thm}

\begin{proof}
We split the proof into four steps. %If $d=1$, then $E_d$ vanishes identically and the assertion holds by always declaring an empty output. We henceforth assume $d\ge2$; in particular, the global menu constructed below is nonempty.

\medskip
\noindent\textbf{Step 1: a countable menu of local robust protocols.}
Let
\[
    \mathsf S\coloneqq \DD(\mathbb C^d\otimes\mathbb C^d)
\]
with the trace-distance topology.  This is a compact metric space, hence it is
second countable.  In particular, every open subset of \(\mathsf S\) is
second countable, and every open cover of an open subset has a countable
subcover.

Fix a positive rational number \(q\).  Define
\[
    \mathcal O_q
    \coloneqq
    \{\gamma\in\mathsf S:E_d(\gamma)>q\}.
\]

For any rational $q$, the set $\mathcal O_q$ is open due to the lower semicontinuity of the distillable entaglement proved in Corollary~\ref{cor:Ed-lower-semicontinuity}.
\begin{comment}
We first check that \(\mathcal O_q\) is open.  Let \(\gamma\in\mathcal O_q\).
Choose a real number \(q^+(\gamma,q)\) such that
\begin{equation}
    q<q^+(\gamma,q)<E_d(\gamma).
\label{eq:auxiliary-rate}
\end{equation}
By the local robustness lemma, applied at rate \(q^+(\gamma,q)\),
there exist a radius \(r(\gamma,q)>0\) and a sequence of LOCC maps
\(\widetilde\Lambda_n^{(\gamma,q)}\) such that
\begin{equation}
    \lim_{n\to\infty}
    \sup_{\sigma\in B_{r(\gamma,q)}(\gamma)}
    \frac12
    \left\|
        \widetilde\Lambda_n^{(\gamma,q)}(\sigma^{\otimes n})
        -
        \PhiTwo^{\otimes\lceil q^+(\gamma,q)n\rceil}
    \right\|_1
    =0.
\label{eq:local-robust-at-auxiliary-rate}
\end{equation}
Therefore every state in \(B_{r(\gamma,q)}(\gamma)\) has distillable
entanglement at least \(q^+(\gamma,q)>q\), and hence belongs to
\(\mathcal O_q\).  Thus \(\mathcal O_q\) is open. Being an open subspace of the second-countable space \(\mathsf S\), it is itself second countable.
\end{comment}

In this first part of the proof we have to define the concept of \textit{menu}, which will be used extensively throughout the proof. Intuitively, the label \(q\) of the menu entry means: \textit{this entry guarantees distillable rate \(q\)}. In order to make this definition rigorous, we start by taking the open balls

\begin{comment} we label the entry by the guaranteed output rate \(q\).
%not by the auxiliary rate \(q^+(\gamma,q)\).  Since
\(q^+(\gamma,q)>q\), for all sufficiently large \(n\) one has
\[
    \lceil q^+(\gamma,q)n\rceil\ge \lceil qn\rceil.
\]
By discarding surplus Bell pairs after \(\widetilde\Lambda_n^{(\gamma,q)}\),
and by defining the finitely many small blocklengths arbitrarily, we obtain a
sequence of LOCC maps \(\Lambda_n^{(\gamma,q)}\) satisfying
\begin{equation}
    \lim_{n\to\infty}
    \sup_{\sigma\in B_{r(\gamma,q)}(\gamma)}
    \frac12
    \left\|
        \Lambda_n^{(\gamma,q)}(\sigma^{\otimes n})
        -
        \PhiTwo^{\otimes\lceil qn\rceil}
    \right\|_1
    =0.
\label{eq:local-robust-at-labelled-rate}
\end{equation}
Thus the label \(q\) of the menu entry means: ``this entry guarantees rate
\(q\)''.   In
fact, the center satisfies \(E_d(\gamma)>q\), and the robustness radius may
have been obtained using the auxiliary rate \(q^+(\gamma,q)>q\).
\end{comment}

\begin{equation}
    B_{r(\gamma,q)/4}(\gamma),
    \qquad \gamma\in\mathcal O_q.
\label{eq:q-cover-before-countable-subcover}
\end{equation}
They form an open cover of \(\mathcal O_q\).  Notice that this is a whole family of
balls for the fixed value of \(q\): there is not one ball attached to \(q\), but
one ball for each \(\gamma\in\mathcal O_q\).

Since \(\mathcal O_q\) is second countable, we can extract a countable subcover.
Thus, for this fixed rational \(q\), there are centers
\(\gamma_{q,j}\in\mathcal O_q\), radii \(r_{q,j}>0\), and LOCC protocols
\(\Lambda_n^{q,j}\), indexed by \(j\in J_q\subseteq\mathbb N\), such that
\begin{equation}
    \mathcal O_q
    =
    \bigcup_{j\in J_q} B_{r_{q,j}/4}(\gamma_{q,j}),
\label{eq:Oq-covered-by-q-menu}
\end{equation}
and, for every fixed \(j\in J_q\),
\begin{equation}
    \lim_{n\to\infty}
    \sup_{\sigma\in B_{r_{q,j}}(\gamma_{q,j})}
    \frac12
    \left\|
        \Lambda_n^{q,j}(\sigma^{\otimes n})
        -
        \PhiTwo^{\otimes\lceil qn\rceil}
    \right\|_1
    =0.
\label{eq:qj-entry-error-vanishes}
\end{equation}
If \(\mathcal O_q=\varnothing\), then \(J_q=\varnothing\) and this rational
\(q\) contributes no entries.

It is useful to name the objects precisely.

\begin{itemize}

    \item A \emph{menu entry} is one quadruple
    \begin{equation}
        \mathsf E(q,j)
        \coloneqq
        \bigl(q,\gamma_{q,j},r_{q,j},\Lambda_n^{q,j}\bigr).
    \label{eq:qj-entry-definition}
    \end{equation}
    It consists of a guaranteed rate, a center, a robustness radius, and a
    sequence of local robust protocols.

    \item The \emph{\(q\)-menu} is the countable family
    \[
        \mathcal M(q)
        \coloneqq
        \{\mathsf E(q,j):j\in J_q\}.
    \]

    \item The \emph{global menu} is the disjoint union of all \(q\)-menus: 
    \begin{equation}
        \mathcal M
        \coloneqq
        \bigsqcup_{q\in\mathbb Q_{>0}}\mathcal M(q)
        =
        \{\mathsf E(q,j):q\in\mathbb Q_{>0},\ j\in J_q\}.
    \label{eq:global-menu-disjoint-union}
    \end{equation}
\end{itemize}

The word ``disjoint'' is important only notationally.  Even if two entries have
the same center, the same radius, or the same protocol, they are still treated
as distinct entries if they come from different labels \((q,j)\).  Since
\(\mathbb Q_{>0}\) is countable and each \(J_q\) is countable, the global menu
\(\mathcal M\) is countable.  Choose once and for all an enumeration of it:
\begin{equation}
    a\in\mathbb N
    \longmapsto
    (q(a),j(a))\in\mathcal M.
\label{eq:global-menu-enumeration}
\end{equation}
For the entry with global index \(a\), define
\begin{equation}
    q_a\coloneqq q(a),
    \qquad
    \gamma_a\coloneqq \gamma_{q(a),j(a)},
    \qquad
    r_a\coloneqq r_{q(a),j(a)},
    \qquad
    \Lambda_n^a\coloneqq \Lambda_n^{q(a),j(a)}.
\label{eq:global-entry-data}
\end{equation}

The map \(a\mapsto q_a\) is not injective, and it should not be thought of as
assigning a single ball to each rational rate.  For a fixed rational \(q\), set
\begin{equation}
    \mathcal I_q
    \coloneqq
    \{a\in\mathbb N:q_a=q\}.
\label{eq:indices-with-rate-q}
\end{equation}
Then \(\mathcal I_q\) is precisely the set of global indices corresponding to
the whole \(q\)-menu.  Therefore \eqref{eq:Oq-covered-by-q-menu} becomes
\begin{equation}
    \mathcal O_q
    =
    \bigcup_{a\in\mathcal I_q}B_{r_a/4}(\gamma_a).
\label{eq:Oq-covered-by-global-indices}
\end{equation}
This is the exact covering property used later in the proof.

Finally, for each global entry \(a\), define its own worst-case error at
blocklength \(m\) by
\begin{equation}
    e_{a,m}
    \coloneqq
    \sup_{\sigma\in B_{r_a}(\gamma_a)}
    \frac12
    \left\|
        \Lambda_m^a(\sigma^{\otimes m})
        -
        \PhiTwo^{\otimes\lceil q_a m\rceil}
    \right\|_1 .
\label{eq:entry-error-global-index}
\end{equation}
For every fixed \(a\), the construction gives
\begin{equation}
    \lim_{m\to\infty}e_{a,m}=0.
\label{eq:entry-error-global-index-vanishes}
\end{equation}

\medskip
\noindent\textbf{Step 2: the activation schedule.}
At block length \(n\), the universal protocol will not use the whole infinite menu. It will activate only the first \(A(n)\) entries. The function \(A(n)\) must tend to infinity, so that every fixed useful entry eventually becomes available. But it must tend to infinity slowly enough that tomography at the required accuracy still costs only \(o(n)\) copies.

For a finite menu size \(M\ge1\), define the accuracy assigned to the first \(M\) entries by
\begin{equation}
    \delta_M
    \coloneqq
    \min\left\{
        \frac12,
        \frac1M,
        \frac{r_1}{4},
        \frac{r_2}{4},
        \ldots,
        \frac{r_M}{4}
    \right\}.
\label{eq:delta-M-definition}
\end{equation}
There are two reasons for this definition. First, \(\delta_M\le r_a/4\) for every active entry \(a\le M\). This is the safety margin needed to guarantee that a protocol selected from the estimate also applies to the true state. Second, \(\delta_M\le1/M\), and therefore
\begin{equation}
    \lim_{M\to\infty}\delta_M = 0.
\label{eq:delta-M-goes-zero}
\end{equation}
%This removes any ambiguity about whether the tomographic accuracy really tends to zero.

Fix the constant \(C_{\rm tomo}\) such that Theorem~\ref{thm:LOCC_tomography_qudit_qudit} can be implemented using at most
\begin{equation}
    \left\lceil
    C_{\rm tomo}
    \frac{d^6+d^2\log(1/p)}{\varepsilon^2}
    \right\rceil
\label{eq:tomo-explicit-bound}
\end{equation}
copies for accuracy \(\varepsilon\) and failure probability \(p\). Hence tomography with accuracy \(\delta_M\) and failure probability \(n^{-2}\) uses at most
\begin{equation}
    t(n,M)
    \coloneqq
    \left\lceil
    C_{\rm tomo}
    \frac{d^6+d^2\log(n^2)}{\delta_M^2}
    \right\rceil
\label{eq:tomo-copies-n-M}
\end{equation}
copies. Here \(M\) is fixed while \(n\) tends to infinity.

For each fixed \(M\), the number \(\delta_M\) is a positive constant. Therefore
\[
    t(n,M)=O(\log n)=o(n)
    \qquad(n\to\infty,\; M\text{ fixed}).
\]
Also, because only finitely many errors \(e_{a,m}\) with \(a\le M\) are involved, \eqref{eq:entry-error-global-index} implies that
\[
    \lim_{m\to\infty} \max_{1\le a\le M}e_{a,m} = 0 .
\]
Consequently, for every \(M\ge1\), we may choose an integer \(N_M\) so large that the following two estimates hold for all \(n\ge N_M\):
\begin{align}
    \max_{1\le a\le M}e_{a,m}
    &\le \frac1M
    \qquad\text{for every }m\ge \frac n2,
\label{eq:NM-error-condition}\\[0.5em]
    t(n,M)&\le \min\left\{\frac n2,\frac{n}{M}\right\}.
\label{eq:NM-tomo-condition}
\end{align}
We choose the thresholds recursively so that
\begin{equation}
    2\le N_1<N_2<N_3<\cdots,
    \qquad
    N_M\ge M.
\label{eq:NM-increasing}
\end{equation}
This recursive choice is possible because the two requirements above are eventually true for each fixed \(M\) as $n$ goes to infinity.

Now the direction of dependence is fixed once and for all:
\[
    M\quad\longmapsto\quad N_M
\]
is chosen first, as part of the design of the universal protocol. After this has been done, the block length \(n\) determines how much of the menu is active. Define
\begin{equation}
    A(n)
    \coloneqq
    \max\{M\ge1:N_M\le n\}
\label{eq:A-n-definition}
\end{equation}
for \(n\ge N_1\), and define \(A(n)=1\) for the finitely many smaller \(n\). Thus, equivalently, if
\[
    N_M\le n<N_{M+1},
\]
then
\[
    A(n)=M.
\]
Since \(N_M\) is a strictly increasing positive sequence, we have
\begin{equation}
    \lim_{n\to\infty} A(n) = \infty.
\label{eq:A-n-goes-infty}
\end{equation}
Finally define the actual tomographic accuracy used at block length \(n\) by the composition
\begin{equation}
    \delta(n)
    \coloneqq
    \delta_{A(n)}.
\label{eq:delta-n-definition}
\end{equation}
This is not just a change of notation: \(\delta_M\) is a sequence indexed by the planned finite menu size \(M\), while \(\delta(n)\) is the accuracy used at the actual block length \(n\).

The two crucial consequences are now immediate. First,
\[
    0<\delta(n)=\delta_{A(n)}\le\frac1{A(n)},
\]
and hence
\begin{equation}
    \lim_{n\to\infty}\delta(n)=0.
\label{eq:delta-n-goes-zero}
\end{equation}
Second, if
\begin{equation}
    m_n
    \coloneqq
    t(n,A(n))
\label{eq:m-n-definition}
\end{equation}
is the number of copies used for tomography at block length \(n\), then \eqref{eq:NM-tomo-condition} with \(M=A(n)\) gives
\begin{equation}
    m_n
    \le
    \frac{n}{A(n)}.
\label{eq:m-n-sublinear-bound}
\end{equation}
Therefore
\begin{equation}
    \lim_{n\to\infty}\frac{m_n}{n}=0.
\label{eq:m-n-sublinear}
\end{equation}
In particular, the number of unused copies left for distillation,
\begin{equation}
    s_n\coloneqq n-m_n,
\label{eq:s-n-definition}
\end{equation}
satisfies
\begin{equation}
    \lim_{n\to\infty}\frac{s_n}{n}=1.
\label{eq:s-n-properties}
\end{equation}
Since \(A(n)\to\infty\), one has \(A(n)\ge2\) for all sufficiently large \(n\). For those \(n\), Eq.~\eqref{eq:m-n-sublinear-bound} gives \(m_n\le n/2\), and hence \(s_n\ge n/2\). We may therefore apply \eqref{eq:NM-error-condition} with \(M=A(n)\) and \(m=s_n\), obtaining, for all sufficiently large \(n\),
\[
    \max_{1\le a\le A(n)}e_{a,s_n}
    \le\frac1{A(n)}.
\]
Consequently,
\begin{equation}
    \lim_{n\to\infty}\max_{1\le a\le A(n)}e_{a,s_n}=0.
\label{eq:active-errors-vanish}
\end{equation}
This is the diagonal estimate: the active menu grows, the required accuracy goes to zero, the tomographic cost is sublinear, and the worst error among active local protocols still vanishes.

\medskip
\noindent\textbf{Step 3: definition of the universal protocol.}
We now define the \(n\)-th protocol. For the finitely many values \(n<N_1\), let $\mathcal U_n$ output the failure label and no Bell-pair registers; the readout $\mathcal D_{n,R}$ then prepares the fixed product state $\ketbra{00}^{\otimes\ceil{Rn}}$. For \(n\ge N_1\), proceed as follows.

First, Alice and Bob use the first \(m_n\) copies for LOCC tomography with accuracy \(\delta(n)\) and failure probability \(n^{-2}\). Since the failure probabilities are summable, we can invoke the Borel--Cantelli lemma to say that the tomographic procedure can fail at most a finite number of times.
%almost surely. 
They obtain a classical estimate \(\widehat\rho_n\). For every true input state \(\rho\), the tomographic success event satisfies
\begin{equation}
    \Pr
    \left[
        \frac12\|\widehat\rho_n-\rho\|_1\le\delta(n)
    \right] \ge 1 - n^{-2}.
\label{eq:tomo-success-event-pointwise}
\end{equation}
The remaining \(s_n=n-m_n\) copies are untouched and are still in the state \(\rho^{\otimes s_n}\).

Second, after observing \(\widehat\rho_n\), Alice and Bob inspect only the active finite menu
\[
    a=1,2,\ldots,A(n).
\]
They call an active entry \(a\) \textit{admissible} if
\begin{equation}
    \widehat\rho_n\in B_{r_a/2}(\gamma_a).
\label{eq:admissible-entry-state}
\end{equation}
If at least one active entry is admissible, they choose one with largest rate \(q_a\). If no active entry is admissible, they choose the failure label \(a=0\). This choice is a classical function of the tomographic transcript, and hence it is part of an LOCC protocol.

Third, for an active entry define
\[
    k_{a,n}\coloneqq\ceil{q_as_n},
    \qquad
    K_n\coloneqq\max_{1\le a\le A(n)}k_{a,n},
\]
and set \(k_{0,n}\coloneqq0\) for the failure label.
The universal map \(\mathcal U_n\) has a fixed output space: a classical register containing the selected label and \(K_n\) shared qubit-pair registers. If the selected label is \(a\ge1\), Alice and Bob apply \(\Lambda_{s_n}^a\), place its \(k_{a,n}\) output pairs in the first \(k_{a,n}\) slots, and fill the remaining slots with the product state \(\ketbra{00}\). On the failure label \(a=0\), they fill all slots with \(\ketbra{00}\). This padding is local, so \(\mathcal U_n\) is an LOCC channel.
%with a branch-independent codomain.

For every target rate \(R>0\), define a deterministic LOCC readout \(\mathcal D_{n,R}\) as follows. It reads the classical label. If the declared length \(k_{a,n}\) is at least \(\ceil{Rn}\), it keeps the first \(\ceil{Rn}\) pair registers and discards all the others. Otherwise, it discards the padded register and locally prepares \(\ketbra{00}^{\otimes\ceil{Rn}}\). The map \(\mathcal D_{n,R}\) depends only on \(n\) and \(R\), not on the unknown input state.

\medskip
\noindent\textbf{Step 4: performance analysis.}
Fix the true state \(\rho\in\mathsf S\) and a target rate
\(0<R<E_d(\rho)\).  Choose a rational number \(q\) such that
\begin{equation}
    R<q<E_d(\rho).
\label{eq:choose-rational-q}
\end{equation}
Then \(\rho\in\mathcal O_q\).  By the covering property of the \(q\)-menu,
written in global indices as \eqref{eq:Oq-covered-by-global-indices}, there is
at least one index
\begin{equation}
    a_*\in\mathcal I_q=\{a:q_a = q\}
\label{eq:a-star-in-q-menu}
\end{equation}
such that
\begin{equation}
    \rho\in B_{r_{a_*}/4}(\gamma_{a_*}).
\label{eq:rho-in-a-star-ball}
\end{equation}
Equivalently,
\begin{equation}
    q_{a_*}=q,
    \qquad
    \rho\in B_{r_{a_*}/4}(\gamma_{a_*}).
\label{eq:good-entry-state-clarified}
\end{equation}

Since the activation function satisfies \(A(n)\to\infty\), the fixed entry
\(a_*\) is active for all sufficiently large \(n\).  On the tomographic success
event, and for such large \(n\), the construction gives
\[
    \delta(n)=\delta_{A(n)}\le \frac{r_{a_*}}4.
\]
Together with \eqref{eq:rho-in-a-star-ball}, this implies
\begin{align}
    \frac12\norm{\widehat\rho_n-\gamma_{a_*}}_1 \nonumber
    &\le \nonumber
    \frac12\norm{\widehat\rho_n-\rho}_1
    +
    \frac12\norm{\rho-\gamma_{a_*}}_1 \\
    &< \nonumber
    \frac{r_{a_*}}4+\frac{r_{a_*}}4 \\
    &=
    \frac{r_{a_*}}2 .
\end{align}
Hence the entry \(a_*\) is admissible.  Therefore, whenever tomography
succeeds, the set of indices that are both \textit{active} and \textit{admissible} is nonempty.  Since the universal protocol
chooses an admissible active entry of largest guaranteed rate, the selected
entry \(c\) satisfies:
\begin{equation}
    q_c\ge q>R.
\label{eq:selected-entry-rate-lower-bound}
\end{equation}

We must also check that the selected local protocol is valid for the true state, not merely for the estimate. Since the selected entry \(c\) is active, \(c\le A(n)\), and hence
\[
    \delta(n)=\delta_{A(n)}\le\frac{r_c}{4}.
\]
Since it is admissible,
\[
    \frac12\|\widehat\rho_n-\gamma_c\|_1<\frac{r_c}{2}.
\]
On the tomography success event,
\[
    \frac12\|\rho-\widehat\rho_n\|_1\le\delta(n)\le\frac{r_c}{4}.
\]
Therefore
\begin{equation}
    \frac12\|\rho-\gamma_c\|_1
    \le
    \frac12\|\rho-\widehat\rho_n\|_1
    +
    \frac12\|\widehat\rho_n-\gamma_c\|_1
    <
    \frac{r_c}{4}+\frac{r_c}{2}
    <r_c.
\label{eq:true-state-in-selected-ball}
\end{equation}
Thus \(\rho\in B_{r_c}(\gamma_c)\), exactly the region on which entry \(c\) was guaranteed to work.

On the tomographic success event, the trace-distance error of the selected local protocol at its declared output length is at most
\[
    e_{c,s_n}
    \le
    \max_{1\le b\le A(n)}e_{b,s_n}.
\]
It remains to compare that declared length with the target rate \(R\). Since \(s_n/n\to1\) and \(q>R\), there exists \(n_0\) such that
\[
    qs_n\ge Rn+1
    \qquad(n\ge n_0).
\]
For all such \(n\), Eq.~\eqref{eq:selected-entry-rate-lower-bound} implies
\[
    k_{c,n}=\ceil{q_cs_n}
    \ge
    \ceil{qs_n}
    \ge
    \ceil{Rn}.
\]
Thus, on the tomographic success event and for all sufficiently large \(n\), the readout \(\mathcal D_{n,R}\) only discards surplus registers. By data processing, the conditional error after this readout is still at most \(\max_{b\le A(n)}e_{b,s_n}\). On the tomographic failure event, whose probability is at most \(n^{-2}\), the trace distance from the target is at most one. Convexity of the trace norm therefore gives
\begin{equation}
    \lim_{n\to\infty}\half
    \norm{
        \mathcal D_{n,R}\!\circ\mathcal U_n(\rho^{\otimes n})
        -\Phi_2^{\otimes\ceil{Rn}}
    }_1
    \le \lim_{n\to\infty} \left(
    n^{-2}
    +
    \max_{1\le b\le A(n)}e_{b,s_n} \right) = 0 .
\label{eq:pointwise-total-error}
\end{equation}

The same safety-margin argument applies to every state, even when $E_d(\rho)=0$. On tomographic success, every selected entry is valid for $\rho$. Thus, the trace-distance error averaged over the declared length is at most 
\begin{equation}
\lim_{n\to\infty}\left(n^{-2}+\max_{b\le A(n)}e_{b,s_n}\right) = 0 .
\end{equation}

%; the empty-output branch has zero error. 
The rate argument above also gives $\Pr[k_{c,n}<\ceil{Rn}]\le n^{-2}$ for all sufficiently large $n$. Since \(\rho\) and \(0<R<E_d(\rho)\) were arbitrary, the same sequence \(\mathcal U_n\) has the claimed universal property. This proves the theorem.
\end{proof}

\section{Proof of Theorem 4 of the main text}

Here, we prove the worst-case performance of a universal protocol given a generic family of states $\XX\subseteq\DD(\mathbb{C}^d\otimes\mathbb{C}^d)$. For every nonempty family \(\XX\) and every \(\varepsilon\in[0,1)\), define
\begin{equation}
\begin{aligned}
    \fament^\varepsilon(\XX)
    \coloneqq
    \sup\Bigg\{R>0:\;&
    \limsup_{n\to\infty}
    \inf_{\Lambda_n\in\locc(A^nB^n\to A_0^{\ceil{Rn}}B_0^{\ceil{Rn}})}
    \sup_{\rho\in\XX}
    \half
    \norm{\Lambda_n(\rho^{\otimes n})-
    \Phi_2^{\otimes\ceil{Rn}}}_1
    \le\varepsilon
    \Bigg\},
\end{aligned}
\label{eq:family-threshold-definition}
\end{equation}
with the convention that the supremum of an empty set of positive rates is zero, and set \(\fament(\XX)\coloneqq\fament^0(\XX)\). Since the errors are nonnegative, the condition at \(\varepsilon=0\) is equivalent to convergence to zero. We first prove a useful lemma.

\begin{lemma}[Stability under closure]
\label{lem:closure_universal}
For every \(\varepsilon\in[0,1)\) and every nonempty family
\(\XX\subseteq\DD(\mathbb C^d\otimes\mathbb C^d)\),
\[
    \fament^\varepsilon(\XX)
    =
    \fament^\varepsilon(\overline{\XX}) .
\]
In particular,
\[
    \fament(\XX)=\fament(\overline{\XX}) .
\]
\end{lemma}

\begin{proof}
Fix \(n\), \(R\), and an LOCC map \(\Lambda_n\). The function
\[
    \rho\longmapsto
    \half
    \norm{\Lambda_n(\rho^{\otimes n})-
    \Phi_2^{\otimes\ceil{Rn}}}_1
\]
is continuous in trace norm. Therefore its supremum over \(\XX\) equals its supremum over \(\overline{\XX}\). This equality holds before taking the infimum over \(\Lambda_n\), before taking the limsup in \(n\), and before taking the supremum over \(R\). Hence the definition of \(\fament^\varepsilon\) is unchanged when \(\XX\) is replaced by \(\overline{\XX}\).
\end{proof}

Now, we proceed to the proof of the last Theorem.

\begin{thm}[Worst-case family distillation]
\label{thm:main_vanishing}
For every nonempty family
\(\XX\subseteq\DD(\mathbb C^d\otimes\mathbb C^d)\),
\begin{equation}
    \fament(\XX)
    =
    \fament(\overline{\XX})
    =
    \min_{\rho\in\overline{\XX}}E_d(\rho) .
\label{eq:main_result}
\end{equation}
In particular, if \(\XX\) is closed, then
\[
    \fament(\XX)=\min_{\rho\in\XX}E_d(\rho).
\]
\end{thm}

\begin{proof}
By Lemma~\ref{lem:closure_universal}, it is enough to prove the claim for closed families. Hence, for the rest of the proof, assume \(\XX\) is closed. Since the state space is compact in finite dimension, \(\XX\) is compact. By Corollary~\ref{cor:Ed-lower-semicontinuity}, \(E_d\) is lower semicontinuous, and therefore it attains its minimum on the nonempty compact set \(\XX\). Thus the infima appearing below are minima.

\medskip
\noindent\textbf{Upper bound.}
Suppose a rate \(R\) is universally achievable with vanishing error on \(\XX\). Then there are LOCC maps \(\Lambda_n\) such that
\[
    \lim_{n\to\infty}
    \sup_{\rho\in\XX}
    \half
    \norm{\Lambda_n(\rho^{\otimes n})-
    \Phi_2^{\otimes\ceil{Rn}}}_1
    =0 .
\]
Fix any \(\rho\in\XX\). The same sequence \(\Lambda_n\) is a valid state-dependent distillation protocol for this particular \(\rho\). Therefore \(R\le E_d(\rho)\). Since this holds for every \(\rho\in\XX\),
\[
    R\le\inf_{\rho\in\XX}E_d(\rho).
\]
Taking the supremum over all universally achievable \(R\) gives
\begin{equation}
    \fament(\XX)
    \le
    \inf_{\rho\in\XX}E_d(\rho).
\label{eq:upper_bound}
\end{equation}

\medskip
\noindent\textbf{Lower bound.}
If \(\inf_{\rho\in\XX}E_d(\rho)=0\), there is nothing to prove. Assume therefore that this infimum is positive, and fix
\[
    0<R<\inf_{\rho\in\XX}E_d(\rho).
\]
Choose an intermediate rate \(R_0\) satisfying
\[
    R<R_0<\inf_{\rho\in\XX}E_d(\rho).
\]
For every \(\rho\in\XX\), Lemma~\ref{lem:local_universal_robustness}, applied with rate \(R_0\), gives a radius \(r_\rho>0\) and a sequence of LOCC maps that works uniformly on \(B_{r_\rho}(\rho)\). The smaller balls
\[
    B_{r_\rho/4}(\rho),
    \qquad \rho\in\XX,
\]
still cover \(\XX\). By compactness, choose a finite subcover:
\begin{equation}
    \XX\subseteq
    \bigcup_{i=1}^M B_{r_i/4}(\rho_i),
\label{eq:finite_cover}
\end{equation}
where \(r_i\) is the robustness radius associated with the center \(\rho_i\). Let
\begin{equation}
    \Delta\coloneqq\frac14\min_{1\le i\le M}r_i>0 .
\label{eq:Delta}
\end{equation}

We now define one universal protocol on \(n\) copies of an unknown state \(\rho\in\XX\).

First, Alice and Bob use
\[
    m_n=O\!\left(\frac{d^6+d^2\log n}{\Delta^2}\right)
\]
copies for LOCC tomography, with accuracy \(\Delta\) and failure probability \(n^{-2}\). Thus
\begin{equation}
    \Pr\!\left[
        \half\norm{\widehat\rho_n-\rho}_1\le\Delta
    \right]
    \ge 1-n^{-2} .
\label{eq:tomo_success}
\end{equation}
For fixed \(d\) and \(\Delta\), we have \(m_n=o(n)\). Set
\[
    s_n\coloneqq n-m_n .
\]
For small \(n\), if \(s_n\le0\), the protocol may output a fixed product state. This does not affect the asymptotic statement. Below we consider large \(n\), for which \(s_n>0\).

Second, after obtaining \(\widehat\rho_n\), Alice and Bob choose the first index \(i\) such that
\[
    \widehat\rho_n\in B_{r_i/2}(\rho_i).
\]
If no such index exists, they choose \(i=1\). This exceptional branch only matters when tomography fails.

Let us check that, on the tomographic success event in \eqref{eq:tomo_success}, the chosen local protocol is valid for the true state. Since \(\rho\in\XX\), the finite cover \eqref{eq:finite_cover} gives an index \(j\) such that
\[
    \rho\in B_{r_j/4}(\rho_j).
\]
Together with \(\half\norm{\widehat\rho_n-\rho}_1\le\Delta\le r_j/4\), this implies
\[
    \widehat\rho_n\in B_{r_j/2}(\rho_j).
\]
Hence at least one admissible index exists. If the protocol chooses index \(i\), then
\[
    \widehat\rho_n\in B_{r_i/2}(\rho_i).
\]
Using again \(\half\norm{\widehat\rho_n-\rho}_1\le\Delta\le r_i/4\), we get
\[
    \half\norm{\rho-\rho_i}_1
    \le
    \half\norm{\rho-\widehat\rho_n}_1
    +
    \half\norm{\widehat\rho_n-\rho_i}_1
    <
    \frac{r_i}{4}+\frac{r_i}{2}
    <r_i .
\]
Therefore \(\rho\in B_{r_i}(\rho_i)\), so the robustness guarantee for center \(\rho_i\) applies to the true state.

Third, Alice and Bob apply the corresponding local robust distillation map to the remaining \(s_n\) copies. That map distills at rate \(R_0\) with error going to zero uniformly over \(B_{r_i}(\rho_i)\). Since \(m_n=o(n)\),
\[
    \lim_{n\to\infty}\frac{s_n}{n} = 1 .
\]
Therefore, for all sufficiently large \(n\),
\[
    \ceil{R_0s_n}\ge\ceil{Rn} .
\]
Let \(D_{i,n}\) be the LOCC channel that discards the surplus Bell pairs after the local protocol chosen by index \(i\). It satisfies
\[
    D_{i,n}\!\left(\Phi_2^{\otimes\ceil{R_0s_n}}\right)
    =
    \Phi_2^{\otimes\ceil{Rn}} .
\]
The global protocol is: run tomography, choose \(i\), apply \(\Lambda_{s_n}^{(\rho_i,R_0)}\) to the remaining copies, and then apply \(D_{i,n}\). This is LOCC because it is a classical conditioning over LOCC branches.

Let
\[
    e_{i,n}
    \coloneqq
    \sup_{\sigma\in B_{r_i}(\rho_i)}
    \half
    \norm{D_{i,n}\!\left(\Lambda_{s_n}^{(\rho_i,R_0)}(\sigma^{\otimes s_n})\right)-
    \Phi_2^{\otimes\ceil{Rn}}}_1 .
\]
By data processing and the defining property of \(\Lambda_{s_n}^{(\rho_i,R_0)}\), each \(e_{i,n}\to0\). Since there are finitely many indices,
\[
    \lim_{n\to\infty}\max_{1\le i\le M}e_{i,n} = 0 .
\]
For any input \(\rho\in\XX\), the tomographic failure event has probability at most \(n^{-2}\). 
%On that event, the trace distance is at most one. 
On the success event, the chosen branch has error at most \(\max_i e_{i,n}\). Hence
\[
\begin{aligned}
\lim_{n\to\infty}\sup_{\rho\in\XX}
\half
\norm{\Lambda_n(\rho^{\otimes n})-
\Phi_2^{\otimes\ceil{Rn}}}_1 
\le
\lim_{n\to\infty} \left(n^{-2}+\max_{1\le i\le M}e_{i,n}\right) = 0 .
\end{aligned}
\]
Thus \(R\) is universally achievable with vanishing error on \(\XX\). Since the proof holds for every
\[
    R<\inf_{\rho\in\XX}E_d(\rho) ,
\]
we obtain
\begin{equation}
    \fament(\XX)
    \ge
    \inf_{\rho\in\XX}E_d(\rho).
\label{eq:lower_bound}
\end{equation}
Combining \eqref{eq:upper_bound} and \eqref{eq:lower_bound} proves the theorem for closed \(\XX\). The general statement follows from Lemma~\ref{lem:closure_universal} by replacing \(\XX\) with \(\overline{\XX}\).
\end{proof}

% REVTeX selects apsrev4-2 and writes the longbibliography controls.
\bibliography{biblio}